\documentstyle[12pt]{article}
\def\gsim{\stackrel{>}{{}_\sim}}
\def\be{\begin{equation}}       
\def\ee{\end{equation}}
\def\bear{\be\begin{array}}      
\def\eear{\end{array}\ee}
\def\bea{\begin{eqnarray}}
\def\eea{\end{eqnarray}}

\def\ie{{\it i.e.}}

\def\half{{\textstyle{1 \over 2}}}

\def\quarter{{\textstyle{1 \over 4}}}

\def\eighth{{\textstyle{1 \over 8}}}
\def\bold#1{\setbox0=\hbox{$#1$}%
     \kern-.025em\copy0\kern-\wd0
     \kern.05em\copy0\kern-\wd0
     \kern-.025em\raise.0433em\box0 }
\begin{document}
\catcode`@=11
\newtoks\@stequation
\def\subequations{\refstepcounter{equation}%
\edef\@savedequation{\the\c@equation}%
  \@stequation=\expandafter{\theequation}
  \edef\@savedtheequation{\the\@stequation}
  \edef\oldtheequation{\theequation}%
  \setcounter{equation}{0}%
  \def\theequation{\oldtheequation\alph{equation}}}
\def\endsubequations{\setcounter{equation}{\@savedequation}%
  \@stequation=\expandafter{\@savedtheequation}%
  \edef\theequation{\the\@stequation}\global\@ignoretrue

\noindent}
\catcode`@=12
\begin{titlepage}

\begin{flushright}
IC/96/252\\
IFIC/97-27\\
SHEP--95--38\\
VAND--TH--94--16\\
hep-ph/9705471\\
April 1997
\end{flushright}

\vspace*{5mm}

\begin{center}
{\Huge Diagonalization of Coupled Scalars}\\
\vspace*{5mm}
{\Huge and its Application to the}\\
\vspace*{5mm}
{\Huge  Supersymmetric Neutral Higgs Sector}\\[15mm]

{\large{\bf Marco Aurelio D\'\i az}\\
\hspace{3cm}\\
{\small Departamento de F\'\i sica Te\'orica, Universidad de Valencia}\\ 
{\small Burjassot, Valencia 46100, Spain}
\hspace{3cm}\\
{\small and}
\hspace{3cm}\\
{\small High Energy Section, International Centre for Theoretical 
Physics}\\ 
{\small P.O. Box 586, Trieste 34100, Italy}} 
\end{center}
\vspace{5mm}
\begin{abstract}

We introduce a momentum dependent mixing angle $\alpha(p^2)$
which allow us to diagonalize at any external momentum $p$ the 
one-loop corrected inverse propagator matrix of two coupled 
scalar fields while keeping the full momentum dependence
in the self energies. We compare this method with more traditional 
techniques applied to the diagonalization of coupled scalars at 
the one-loop level. This method is applied to the CP-even Higgs 
sector of the Minimal Supersymmetric Model, defining the momentum 
dependent mixing angle $\alpha(p^2)$, and calculating the two 
CP-even Higgs masses and the mixing angle at these two scales. We 
compare the results obtained in this way with alternative 
techniques. We make explicit the relation between $\alpha(p^2)$ 
and the running mixing angle. We find differences between the 
mixing angle calculated with our method and the one calculated 
with more traditional methods, and these differences are relevant 
for Higgs searches at LEP2.

\end{abstract}

\end{titlepage}

\setcounter{page}{1}

\section{Introduction}

In field theory it is common to find mixing between
different species of scalars, fermions, or vector bosons. In the Standard
Model (SM) we have the mixing between the gauge fields $B$ and $W^3$ 
corresponding to the groups $U(1)$ and $SU(2)$ respectively. After a 
rotation given by the weak mixing angle $\tan\theta_W=g'/g$ at tree
level we find the mass eigenstates $\gamma$ and $Z$. 
Similarly, mixing between
scalar particles are typical of two Higgs doublets models, and in
supersymmetric theories mixing between fermions are also common
(charginos, neutralinos).

In all cases it is trivial to find the mass eigenstates at tree level,
however, one-loop radiative corrections will mix the tree-level 
diagonalized states (for example, $Z$ and $\gamma$ mix at one
loop with charged particles in the loop). Moreover, the sum of the 
mixing graphs will be infinite and momentum dependent. In order to remove 
the infinities, mass counterterms and wave function renormalization 
constants are introduced.

Several years ago, Capdequi-Peyran\`ere and Talon \cite{michael}
studied the wave function renormalization of coupled systems of 
scalars, fermions, and vectors. Their approach include conventional
mass counterterms and wave function renormalization plus a 
non-conventional field ``rotation'' that allow them to
impose no mixing between the states at different scales. These scales
are the masses of the different states. In this paper we
generalize this idea and, at the same time, explain the nature of this
field ``rotation'' by introducing a momentum dependent mixing
angle between two coupled scalars. 

In the Minimal Supersymmetric Model (MSSM) the CP-even Higgs sector
consists of two coupled scalars $H$ and $h$, whose masses satisfy
$m_H>m_h$. Tree level mass relations are simple and specified by
two unknown parameters: the mass of the CP-odd Higgs $m_A$ and
the ratio of the two vacuum expectation value of the two Higgs
doublets $\tan\beta$. Nevertheless, radiative corrections
strongly modify the tree level mass relations, and since the
lightest Higgs might be the first particle detected in the
Higgs sector, it is necessary to achieve a good understanding of the
effect of radiative corrections on this system of coupled scalars.
For this reason, the CP-even sector of the MSSM is the best
place to apply the method of diagonalization of coupled scalars
we are proposing here.

\section{Renormalization of Coupled Scalars}

\subsection{Conventional Wave Function Renormalization}

Similarly to ref.~\cite{michael} (the only difference is that they 
have $m_{12b}^2\equiv0$), consider the bare lagrangian corresponding 
to a system of two scalars:
\begin{equation}
{\cal L}_b=\half\chi_{1b}(p^2-m_{1b}^2)\chi_{1b}+\half\chi_{2b}(p^2-
m_{2b}^2)\chi_{2b}-\chi_{1b}m_{12b}^2\chi_{2b}
\label{eq:Lbare}
\end{equation}
If we denote by $-iA_{ij}^{\chi}(p^2)$, $i,j=1,2$, the sum of the 
one-loop Feynman graphs contributing to the two point 
functions and, after shifting the bare masses by $m_{ib}^2\rightarrow 
m_i^2-\delta m_i^2$, $i=1,2,12$, and the fields by $\chi_{ib}\rightarrow
(1+\half\delta Z_i)\chi_i$, the effective lagrangian is
\begin{equation}
{\cal L}_{eff}=\half(\chi_1,\chi_2){\bold{\Sigma^{\chi}}}
{\chi_1\choose\chi_2}
\label{eq:Leffect}
\end{equation}
where $\bold{\Sigma^{\chi}}$ is the inverse propagator matrix,
with matrix elements given by
\begin{eqnarray}
\Sigma_{11}^{\chi}(p^2)&=&p^2-m_1^2+(p^2-m_1^2)
\delta Z_1+\delta m_1^2-A_{11}^{\chi}(p^2)\nonumber\\
\Sigma_{22}^{\chi}(p^2)&=&p^2-m_2^2+(p^2-m_2^2)
\delta Z_2+\delta m_2^2-A_{22}^{\chi}(p^2)
\label{eq:sigmachielem}\\
\Sigma_{12}^{\chi}(p^2)&=&-m_{12}^2-\half m_{12}^2(\delta Z_1+\delta Z_2)
+\delta m_{12}^2-A_{12}^{\chi}(p^2)\nonumber
\end{eqnarray}
Although it is not a necessary assumption, in order to compare more
easily with ref.~\cite{michael}, we will assume for the moment that the two 
scalars are decoupled at tree level, \ie, $m_{12}^2=0$. In this case,
if we want $m_i$, $i=1,2$, to be the physical masses (the pole of the 
propagators) then the two mass
counterterms are fixed through the relations
\begin{equation}
\delta m_1^2=A_{11}^{\chi}(m_1^2),\qquad\delta m_2^2=A_{22}
^{\chi}(m_2^2)
\label{eq:masscount}
\end{equation}
Similarly, we may want to fix the wave function renormalization constants
by setting to one the residue of each propagator at its pole. In this case
we find
\begin{equation}
\delta Z_1={A'}_{11}^{\chi}(m_1^2),\qquad\delta Z_2={A'}_{22}^{\chi}
(m_2^2)
\label{eq:wavefuncren}
\end{equation}
where the prime denote the derivative with respect to the argument.
We may want to fix the $\delta m_{12}^2$ counterterm by imposing
no mixing between $\chi_1$ and $\chi_2$ at a given scale, for example
at $p^2=m_1^2$. In this case, the off diagonal element of the inverse
propagator matrix is
\begin{equation}
\Sigma_{12}^{\chi}(p^2)=\delta m_{12}^2-A_{12}^{\chi}(p^2)=
A_{12}^{\chi}(m_1^2)-A_{12}^{\chi}(p^2)\equiv-\widetilde A_{12}^{\chi}
(p^2)
\label{eq:sigmaoffdiag}
\end{equation}
At this point all the counterterms are fixed, and since 
$\widetilde A_{12}^{\chi}(p^2)$ is zero only at $p^2=m_1^2$, 
the two fields are not decoupled at a different scale. 
In particular, at the scale given by the mass of the second 
scalar, $p^2=m_2^2$, the mixing is not zero:
\begin{equation}
\Sigma^{\chi}_{12}(m_2^2)=A_{12}^{\chi}(m_1^2)
-A_{12}^{\chi}(m_2^2)\ne0
\label{eq:sigmam2}
\end{equation}
Of course, this non-zero mixing is of one-loop order, and 
if the calculation of the one-loop scalar masses is done 
by diagonalizing the inverse propagator matrix order by order
in perturbation theory, then this non-zero mixing would 
be a two-loop order
effect. Nevertheless, this perturbative diagonalization
can introduce large errors if
radiative corrections are large. And this is the case with
the CP-even Higgs sector of the Minimal 
Supersymmetric Model (MSSM).

\subsection{Mixed Wave Function Renormalization}

According to the previous section, the conventional wave 
function renormalization
gives an inverse propagator matrix which is diagonal only
at one particular scale. This
is the motivation for the authors in ref.~\cite{michael} to define 
the following wave function renormalization $\chi_{1b}\rightarrow
(1-\alpha_1)\chi_1-\beta_1\chi_2$ and $\chi_{2b}\rightarrow(1-\alpha_2)
\chi_2-\beta_2\chi_1$ instead of $\chi_{1b}\rightarrow(1+\half\delta
Z_i)\chi_i$ we use here. If we perform this transformation in the
bare lagrangian in eq.~(\ref{eq:Lbare}) (and taking $m_{12b}^2\equiv0$ 
in order
to follow ref.~\cite{michael}) we find the following inverse propagator
matrix elements 
\begin{eqnarray}
\bar\Sigma_{11}^{\chi}(p^2)&=&p^2-m_1^2-2(p^2-m_1^2)
\alpha_1+\delta m_1^2-A_{11}^{\chi}(p^2)\nonumber\\
\bar\Sigma_{22}^{\chi}(p^2)&=&p^2-m_2^2-2(p^2-m_2^2)
\alpha_2+\delta m_2^2-A_{22}^{\chi}(p^2)
\label{eq:barsigchielem}\\
\bar\Sigma_{12}^{\chi}(p^2)&=&-\beta_1(p^2-m_1^2)-\beta_2(p^2-m_2^2)
-A_{12}^{\chi}(p^2)\nonumber
\end{eqnarray}
where the bar in $\bar\Sigma $is to differentiate with the matrix elements 
in eq.~(\ref{eq:sigmachielem}). Setting to one the residue of the pole of
each propagator they find
\begin{equation}
\alpha_i=-\half{A'}_{ii}^{\chi}(m_1^2)=-\half\delta Z_i,\qquad i=1,2
\label{eq:alfas}
\end{equation}
where we have also included the relation between their $\alpha_i$
and our $\delta Z_i$.
Imposing no mixing between the two fields at $p^2=m_1^2$ and {\it 
also} at $p^2=m_2^2$ they get
\begin{equation}
\beta_1={{A_{12}^{\chi}(m_2^2)}\over{m_1^2-m_2^2}},\qquad\beta_2=
{{A_{12}^{\chi}(m_1^2)}\over{m_2^2-m_1^2}}
\label{eq:betas}
\end{equation}
This procedure is equivalent to take the inverse propagator
matrix in eq.~(\ref{eq:sigmachielem}) in the special case where 
$m_{12}^2=0$ and $\delta m_{12}^2=0$, and 
perform a ``rotation'' (it is not a field rotation in the usual sense,
it is just a wave function renormalization that mixes the two fields)
in the following way
\begin{equation}
{\bold{\Sigma^{\chi}}}\longrightarrow\left[\matrix{1&\beta_2\cr
\beta_1&1\cr}\right]{\bold{\Sigma^{\chi}}}\left[\matrix{1&\beta_1\cr
\beta_2&1\cr}\right]
\label{eq:betarotation}
\end{equation}
with the $\beta_i$ being divergent, as it can be seen from 
eq.~(\ref{eq:betas}). In this case, the inverse propagator matrix in 
eq.~(\ref{eq:barsigchielem})
is simultaneously diagonal at the two different scales $p^2=m_1^2$
and $p^2=m_2^2$. Nevertheless, there will be a non-zero mixing
between the two scalars at any other scale.

It will be instructive to modify the calculation done in 
ref.~\cite{michael} just presented in this section by considering
$m_{12}^2=0$ but $\delta m_{12}^2\ne0$. In this case, the only
modification to the inverse propagator matrix is in the
off-diagonal element in eq.~(\ref{eq:barsigchielem}), and now is equal to
\begin{equation}
\bar\Sigma_{12}^{\chi}(p^2)=-\beta_1^f(p^2-m_1^2)-\beta_2^f
(p^2-m_2^2)+\delta m_{12}^2-A_{12}^{\chi}(p^2)
\label{eq:newbarsig}
\end{equation}
The new coefficients $\beta_i^f$ can be calculated in the same 
way as before, that is imposing no mixing at the scale
$p^2=m_1^2$ and $p^2=m_2^2$. We get
\begin{equation}
\beta_1^f={{A_{12}^{\chi}(m_2^2)-\delta m_{12}^2}\over
{m_1^2-m_2^2}},\quad
\beta_2^f={{A_{12}^{\chi}(m_1^2)-\delta m_{12}^2}\over
{m_2^2-m_1^2}}
\label{eq:betasf}
\end{equation}
The freedom introduced by the new counterterm $\delta m_{12}^2$ 
give us the opportunity to cancel the divergency present in
$A_{12}^{\chi}(p^2)$. This explains the superscript "$f$" in
the constants $\beta_i^f$: they are finite. In this way,
the numerators in eq.~(\ref{eq:betasf}) are just the renormalized
two point function $\widetilde A_{12}^{\chi}(p^2)$ evaluated at
two different scales. The $\beta_i^f$ are then
\begin{equation}
\beta_1^f={{\widetilde A_{12}^{\chi}(m_2^2)}\over
{m_1^2-m_2^2}},\quad
\beta_2^f={{\widetilde A_{12}^{\chi}(m_1^2)}\over
{m_2^2-m_1^2}}
\label{eq:betasftil}
\end{equation}
This time, this wave function renormalization is equivalent
to take the inverse propagator matrix in eq.~(\ref{eq:sigmachielem})
and perform the following ``rotation'':
\begin{equation}
{\bold{\Sigma^{\chi}}}\longrightarrow\left[\matrix{1&\beta^f_2\cr
\beta^f_1&1\cr}\right]{\bold{\Sigma^{\chi}}}\left[\matrix{1&\beta^f_1\cr
\beta^f_2&1\cr}\right]
\label{eq:betarotf}
\end{equation}
Again, the rotated inverse propagator matrix is diagonal only at the 
scales $p^2=m_1^2$ and $p^2=m_2^2$.

\subsection{Momentum Dependent Mixing Angle}

In this paper we introduce a momentum dependent mixing angle $\alpha(p^2)$
which will allow us to diagonalize the inverse propagator matrix at 
{\it any} momentum \cite{thesis,alphadpf}. Considering 
the already finite inverse propagator matrix elements in 
eq.~(\ref{eq:sigmachielem})
(now we work on the general case $m_{12}\ne0$ and $\delta m_{12}^2\ne0$),
we define the momentum dependent mixing angle $\alpha(p^2)$ by
\begin{eqnarray}
\sin 2\alpha(p^2)&=&-{{2\Sigma_{12}^{\chi}(p^2)}\over{\sqrt{
\bigl[\Sigma_{11}^{\chi}(p^2)-\Sigma_{22}^{\chi}(p^2)\bigr]
^2+4\big[\Sigma_{12}^{\chi}(p^2)\big]^2}}}\nonumber\\
\label{eq:defalpha}\\
\cos 2\alpha(p^2)&=&-{{\Sigma_{11}^{\chi}(p^2)-\Sigma_{22}^{\chi}
(p^2)}\over{\sqrt{\bigl[\Sigma_{11}^{\chi}(p^2)-\Sigma_{22}^
{\chi}(p^2)\bigr]^2+4\big[\Sigma_{12}^{\chi}(p^2)\big]^2}}}\nonumber
\end{eqnarray}
The matrix ${\bold{\Sigma^{\chi}}}(p^2)$ is diagonalized at
any momentum $p^2$ by a rotation defined by the angle
$\alpha(p^2)$
\begin{equation}
{\bold{\Sigma^{\chi}}}\longrightarrow\left[\matrix{
c_{\alpha}(p^2)&s_{\alpha}(p^2)\cr-s_{\alpha}(p^2)&
c_{\alpha}(p^2)\cr}\right]{\bold{\Sigma^{\chi}}}
\left[\matrix{c_{\alpha}(p^2)&-s_{\alpha}(p^2)\cr 
s_{\alpha}(p^2)&c_{\alpha}(p^2)\cr}\right]
\label{eq:rotsigma}
\end{equation}
where $s_{\alpha}$ and $c_{\alpha}$ are sine and cosine of the momentum
dependent mixing angle $\alpha(p^2)$. 

In order to find the connection between the mixed wave function 
renormalization introduced in ref.~\cite{michael} and the momentum
dependent mixing angle introduced here, 
consider eqs.~(\ref{eq:sigmachielem}) and (\ref{eq:defalpha}). 
In the case where 
$m_{12}^2=0$ we find in first approximation
\begin{equation}
s_{\alpha}(p^2)\approx{{\widetilde A_{12}^{\chi}(p^2)}\over{m_2^2-m_1^2}},
\qquad c_{\alpha}\approx1
\label{eq:salphaappro}
\end{equation}
where $\widetilde A_{12}^{\chi}(p^2)$ is defined in 
eq.~(\ref{eq:sigmaoffdiag}).
Therefore, the mixed wave function renormalization defined by 
eq.~(\ref{eq:betarotf}) can be obtained from the rotation 
by an angle $\alpha(p^2)$ defined in eq.~(\ref{eq:rotsigma}) 
if we approximate eq.~(\ref{eq:defalpha}) 
to one-loop, order by order in perturbation theory,
and evaluate the non-diagonal entries in the first rotation
matrix in eq.~(\ref{eq:rotsigma}) at two different scales: 
$s_{\alpha}$ at $p^2=m_1^2$ and $-s_{\alpha}$ at $p^2=m_2^2$.

The use of a momentum dependent mixing angle is an alternative
to define a counterterm for this angle. In fact, the renormalization
procedure is carried out in the unrotated basis and no mixing angle is 
defined at that level. Similarly, instead of renormalizing couplings
of the rotated fields to other particles, we renormalize couplings
of the unrotated fields to those particles and after that we rotate
by an angle $\alpha(p^2)$, where $p^2$ is the typical scale of the process,
for example, $p^2=m_i^2$ if the rotated field $\chi_i$ is on-shell.
Usually, working out the radiative corrections in the unrotated basis
implies one extra advantage, and that is the simplicity of the 
Feynman rules. In the following we will illustrate these ideas by
renormalizing the CP-even neutral Higgs masses of the Minimal
Supersymmetric Model (MSSM).

\section{The Minimal Supersymmetric Model}

The radiative corrections to the Higgs masses in the MSSM have 
been studied by many authors in the last few years, 
using three different techniques: the renormalization group 
equation (RGE) method, the effective potential, and the 
diagramatic technique. It was established the
convenience of the parametrization of the Higgs sector with
the CP-odd Higgs mass $m_A$, and the ratio of the
vacuum expectation values of the two Higgs doublets $\tan\beta
=v_2/v_1$. The radiative corrections to the charged Higgs mass
were found to be small \cite{ChargedH,diazhaberi,chaneu},
growing as $m_t^2$, unless there is an appreciable mixing in 
the squark mass matrix: in that case a term proportional to 
$m_t^4$ is non-negligible \cite{diazhaberi}.
The corrections to the CP-even Higgs masses are large and
grow as $m_t^4$, and have profound consequences in the phenomenology
of the Higgs sector \cite{chaneu,neutral,brignole,diazhaberii,yamada}. 
Two-loop corrections
also have been calculated and shown to be 
important \cite{twoloop}.

The MSSM has two Higgs doublets \cite{hhg}:
\begin{equation}
H_1={{\textstyle{1\over{\sqrt{2}}}(\chi_1^0+v_1+i\varphi_1^0)}\choose
{H_1^-}}\qquad H_2={{H_2^+}\choose{\textstyle{1\over{\sqrt{2}}}(\chi_2^0
+v_2+i\varphi_2^0)}}\,,
\label{eq:shift}
\end{equation}
with $v_i$ being the vacuum expectation value of the Higgs fields
$H_i$. In the CP-even Higgs sector there are two
coupled scalar fields $(\chi_1^0,\chi_2^0)$. The
radiatively corrected propagators of the CP-even  
Higgs bosons depend on the two point functions $A_{ij}^{\chi}
(p^2)$ ($i,j=1,2$), where $p^2$ is the external momentum squared.
In perturbative diagonalization of the one-loop mass matrix it is
consistent to consider $p^2$ as a constant equal to the tree level
mass matrix element. Nevertheless, it has been shown that for large
values of the top quark mass, the perturbative diagonalization
of the mass matrix is not reliable. We will
study the effect of those approximations, namely the perturbative
diagonalization of the mass matrix and the replacement of the
external momentum in the self-energies by a constant, and compare
them with the use of the momentum dependent mixing angle $\alpha(p^2)$
that diagonalizes non-perturbatively the inverse propagator
matrix of the CP-even Higgs sector, and we find numerically the
pole of the propagators keeping the full momentum dependence.

\subsection{The model at tree level}
 
We start reviewing the neutral Higgs sector of the MSSM in
the tree level approximation.
The CP-odd mass matrix given by:
\begin{equation}
{\bold{M^2_{\varphi}}}=\pmatrix{m_{12}^2t_{\beta}+t_1/v_1&
m_{12}^2\cr m_{12}^2&m_{12}^2/t_{\beta}+t_2/v_2\cr}
\label{eq:cpoddmasmat}
\end{equation}
where $m_{12}^2$ is a soft symmetry breaking term, 
$t_{\beta}=\tan\beta=v_2/v_1$, and
$t_i$ ($i=1,2$) are the tree level tadpoles, whose expressions
are
\begin{eqnarray}
t_1&=&m_{1H}^2 v_1-m_{12}^2v_2+\textstyle{1\over 8}
(g^2+g'^2)v_1(v_1^2-v_2^2)\,,\nonumber\\
t_2&=&m_{2H}^2 v_2-m_{12}^2 v_1+\textstyle{1\over 8}(g^2+g'^2)
v_2(v_2^2-v_1^2)\,.
\label{eq:tadpoledos}
\end{eqnarray}
Here, $m_{iH}^2=m_i^2+|\mu|^2$ ($i=1,2$), $m_i^2$ are soft
supersymmetry breaking terms, and $\mu$ is the mass parameter
in the superpotential. At tree level and
at the minimum ($t_1=t_2=0$), this matrix is diagonalized with a
rotation of an angle $\beta$. The tree level CP-odd Higgs mass is:
\begin{equation}
(m_A^2)_0={{m_{12}^2}\over{s_{\beta}c_{\beta}}}\,.
\label{eq:cpoddtreemass}
\end{equation}
where the subscript ``0'' refers to the tree level. 
On the other hand, the mass
matrix of the neutral CP-even Higgs bosons is
\begin{equation}
{\bold{M^2_{\chi}}}=\pmatrix{{\textstyle{1\over 4}}
g_Z^2v^2c_{\beta}^2+m_{12}^2t_{\beta}+t_1/v_1&
-{\textstyle{1\over 4}}g_Z^2v^2s_{\beta}c_{\beta}-m_{12}^2
\cr-{\textstyle{1\over 4}}g_Z^2v^2s_{\beta}c_{\beta}-m_{12}^2
&{\textstyle{1\over 4}}g_Z^2v^2s_{\beta}^2+m_{12}^2/t_{\beta}
+t_2/v_2\cr}\,,
\label{eq:cpevmasmat}
\end{equation}
where $g_Z^2=g^2+g'^2$ and $v^2=v_1^2+v_2^2$. From now on, 
we will use the following notation for its matrix elements:
$({\bold{M^2_{\chi}}})_{ij}=m_{ij}^{\chi 2}$.
The tree level masses are obtained setting the tadpoles to
zero and eliminating $m_{12}^2$ in favor of $m_A^2$ using
eq.~(\ref{eq:cpoddtreemass}). The answer is
\begin{equation}
(m^2_{H,h})_0=\half(m_A^2+m_Z^2)\pm
\half\sqrt{(m_Z^2-m_A^2)^2c_{2\beta}^2
+(m_A^2+m_Z^2)^2s_{2\beta}^2}\,,
\label{eq:Hhtreemass}
\end{equation}
with a tree level mixing angle
\begin{equation}
(\tan 2\alpha)_0={{(m_A^2+m_Z^2)}\over{
(m_A^2-m_Z^2)}}\tan 2\beta\,.
\label{eq:alphatree}
\end{equation}
where we have omitted the subscript ``0'' from $m_A^2$.
Next, we calculate the radiative corrections to this 
mixing angle $\alpha$, and introduce the momentum dependent 
mixing angle $\alpha(p^2)$. 

\subsection{One-loop corrections to the CP-even Higgs masses}
 
To find the one loop corrections to the Higgs masses and mixing
angle, we take the mass matrix for the CP-even fields
and replace the bare quantities as indicated by:
\begin{eqnarray}
&H_i\longrightarrow(Z_i)^{1\over 2}H_i\approx(1+\half
\delta Z_i)H_i,\qquad& i=1,2\nonumber\\
&\lambda\longrightarrow \lambda-\delta\lambda,&
\lambda=g,g',e...\nonumber\\
&m\longrightarrow m-\delta m,& m=m_W...,m_{12},...
\label{eq:counter}\\
&v_i\longrightarrow v_i-\delta v_i,& i=1,2.\nonumber
\end{eqnarray}
The one loop effective lagrangian in the CP-even Higgs sector
(we have not rotated the original fields) has the form:
\begin{equation}
{\cal L}=\half(\chi_1^0,\chi_2^0){\bold{\Sigma^{\chi}}}
{\chi_1^0\choose\chi_2^0}
\label{eq:twopointcpev}
\end{equation}
with the matrix elements given by
\begin{eqnarray}
\Sigma_{11}^{\chi}(p^2)&=&\left[p^2-m_{11}^{\chi 2}+
\delta m_{11}^{\chi 2}-A_{11}^{\chi}(p^2)\right]Z_1\nonumber\\
\Sigma_{22}^{\chi}(p^2)&=&\left[p^2-m_{22}^{\chi 2}+
\delta m_{22}^{\chi 2}-A_{22}^{\chi}(p^2)\right]Z_2
\label{eq:sigmaelemZ}\\
\Sigma_{12}^{\chi}(p^2)&=&\left[-m_{12}^{\chi 2}+\delta m_{12}^{\chi 2}
-A_{12}^{\chi}(p^2)\right]Z_1^{1/2}Z_2^{1/2}\nonumber
\end{eqnarray}
where we are allowed to set $t_i=0$ in 
$m_{ii}^{\chi 2}$, given in eq.~(\ref{eq:cpevmasmat}). 
The mass counterterms are:
\begin{eqnarray}
\delta m_{11}^{\chi 2}&=&c_{\beta}^2\delta m_Z^2+s_{\beta}^2
\delta\biggl({{m_{12}^2}\over{s_{\beta}c_{\beta}}}\biggr)
+2(m_A^2-m_Z^2)s_{\beta}^2c_{\beta}^2{{\delta t_{\beta}}\over
{t_{\beta}}}+{{\delta t_1}\over{v_1}}\nonumber\\
\delta m_{22}^{\chi 2}&=&s_{\beta}^2\delta m_Z^2+c_{\beta}^2
\delta\biggl({{m_{12}^2}\over{s_{\beta}c_{\beta}}}\biggr)
-2(m_A^2-m_Z^2)s_{\beta}^2c_{\beta}^2{{\delta t_{\beta}}\over
{t_{\beta}}}+{{\delta t_2}\over{v_2}}
\label{eq:masscountchi}\\
\delta m_{12}^{\chi 2}&=&-s_{\beta}c_{\beta}\delta m_Z^2-s_{\beta}
c_{\beta}\delta\biggl({{m_{12}^2}\over{s_{\beta}c_{\beta}}}\biggr)
-(m_A^2+m_Z^2)s_{\beta}c_{\beta}(c_{\beta}^2-s_{\beta}^2)
{{\delta t_{\beta}}\over{t_{\beta}}}\nonumber
\end{eqnarray}
and to fix them we adopt an on-shell scheme.
Consider first the quadratic terms
in the CP-odd Higgs sector (after rotating by an angle $\beta$):
\begin{equation}
{\cal L}=\half(A,G){\bold{\Sigma_{\varphi}}}{A\choose G}
\label{eq:twopointcpodd}
\end{equation}
where the relevant matrix element is
\begin{eqnarray}
\Sigma_{\varphi}^{AA}(p^2)&=&\left[p^2-m_A^2+\delta m_A^2-A_{AA}(p^2)
\right]Z_A\label{eq:sigmaAGelement}\\
&\approx& p^2-m_A^2+(p^2-m_A^2)\delta Z_A+\delta m_A^2-A_{AA}(p^2)\,.
\nonumber
\end{eqnarray}
and the mass counterterm is given by 
\begin{equation}
\delta m_A^2=\delta\left({{m_{12}^2}\over{s_{\beta}c_{\beta}}}
\right)+s_{\beta}^2{{\delta t_1}\over{v_1}}+c_{\beta}^2{{\delta
t_2}\over{v_2}}
\label{eq:mcountcpodd}
\end{equation}
and the wave function renormalization constant is
\begin{equation}
Z_A=s_{\beta}^2Z_1+c_{\beta}^2Z_2\quad\iff\quad
\delta Z_A=s_{\beta}^2\delta Z_{H_1}+c_{\beta}^2\delta Z_{H_2}\,.
\label{eq:wavefcpodd}
\end{equation}
To fix the mass counterterm $\delta m_A^2$ we
adopt the following on-shell renormalization condition:
\begin{equation}
{\rm Re} \Sigma_{\varphi}^{AA}(m_A^2)=0\quad\Rightarrow\quad
\delta m_A^2={\rm Re} A_{AA}(m_A^2)
\label{eq:Arencond}
\end{equation}
which means that the parameter $m_A^2$ has been defined as the 
pole of the propagator. 

%

Similarly to the previous case, the mass counterterms for the 
$Z$-- and $W$--boson are fixed in an on-shell scheme:
\begin{equation}
\delta m_Z^2={\rm Re}A_{ZZ}(m_Z^2)\,,\qquad 
\delta m_W^2={\rm Re}A_{WW}(m_W^2)\,,
\label{eq:deltamZW}
\end{equation}
\ie, the parameters $m_Z$ and $m_W$ are the pole masses.

Tadpoles are zero at tree level ($t_i=0$), and to fix their
counterterms it is convenient to impose vanishing tadpoles at one-loop
as well. If we call $-i\Gamma_{i}^{(1)}$ to the sum of all Feynman 
diagrams (the 1-point irreducible Green's function)
contributing to the one-loop tadpole, then the tadpole counterterm
is equal to the sum of all one-loop tadpole graphs
\begin{equation}
\delta t_i=\Gamma_i^{(1)}(p^2=0)\,,\qquad i=1,2.
\label{eq:deltatadpole}
\end{equation}

%
%

Now we calculate the counterterm $\delta t_{\beta}$. We define the 
parameter $\tan\beta$ through the $m_{\tilde e_L}$ and $m_{\tilde\nu_e}$
masses \cite{tbdefselec}
\footnote{For alternative definitions of $\tan\beta$ see for example
\cite{diazHdecay,yamadaTB}}.
\begin{equation}
{{\delta t_{\beta}}\over{t_{\beta}}}={1\over{4s_{\beta}^2c_{\beta}^2
m_W^2}}\left[c_{2\beta}A_{WW}(m_W^2)+A_{\tilde e_L\tilde e_L}
(m_{\tilde e_L}^2)-A_{\tilde\nu_e\tilde\nu_e}(m_{\tilde\nu_e}^2)
\right]\,,
\label{eq:tanbetacount}
\end{equation}
where the last two self energies receive no contributions from loops
containing top and bottom quarks and squarks.

In this way, with eqs.~(\ref{eq:mcountcpodd}), (\ref{eq:Arencond}),
(\ref{eq:deltamZW}), (\ref{eq:deltatadpole}), and (\ref{eq:tanbetacount})
we can calculate the mass counterterms in eq.~(\ref{eq:masscountchi}),
which have been fixed in an on--shell scheme where $m_Z$, $m_W$, and $m_A$
are pole masses, and the tadpoles are equal to zero at the one--loop
level.

The pole masses of the two neutral Higgs bosons are the zeros of the
inverse propagator matrix ${\bold{\Sigma^{\chi}}}(p^2)$, whose matrix
elements are defined in eq.~(\ref{eq:sigmaelemZ}), therefore the Higgs 
masses satisfy the following relation
\begin{equation}
\widehat\Sigma_{11}^{\chi}(m_i^2)\widehat\Sigma_{22}^{\chi}(m_i^2)=
\left[\widehat\Sigma_{12}^{\chi}(m_i^2)\right]^2,\qquad i=h,H
\label{eq:detsigmazero}
\end{equation}
where we call
\begin{eqnarray}
\widehat\Sigma_{11}^{\chi}(p^2)&=&p^2-m_Z^2c_{\beta}^2-m_A^2s_{\beta}^2+
\delta m_{11}^{\chi 2}-A_{11}^{\chi}(p^2)\nonumber\\
\widehat\Sigma_{22}^{\chi}(p^2)&=&p^2-m_Z^2s_{\beta}^2-m_A^2c_{\beta}^2+
\delta m_{22}^{\chi 2}-A_{22}^{\chi}(p^2)
\label{eq:sigmaelemnew}\\
\widehat\Sigma_{12}^{\chi}(p^2)&=&(m_Z^2+m_A^2)s_{\beta}c_{\beta}+
\delta m_{12}^{\chi 2}-A_{12}^{\chi}(p^2)\nonumber
\end{eqnarray}
{}From here we see that the wave function renormalization constants
$Z_i$, $i=h,H$ are canceled from the eigenvalue equation, 
implying the independence of the Higgs masses on 
the $Z_i$. In ref. \cite{ChPokorskiR} the Higgs masses are calculated
in this way, although without defining a momentum dependent mixing 
angle. 

The eigenvalues of the matrix ${\bold{\Sigma^{\chi}}}(p^2)$
are the inverse propagators of the two Higgs bosons, and they are
given by
\begin{equation}
\Sigma_i^{\chi}(p^2)=\half\bigl[\Sigma_{11}^{\chi}(p^2)+\Sigma_{22}
^{\chi}(p^2)\bigr]\pm\half\sqrt{\bigl[\Sigma_{11}^{\chi}(p^2)-
\Sigma_{22}^{\chi}(p^2)\bigr]^2+4\big[\Sigma_{12}^{\chi}(p^2)\big]^2}
\label{eq:eigenvalues}
\end{equation}
where $i=H,h$ and the $+$ ($-$) sign correspond to the field $h$ ($H$).
These inverse propagators have a zero at the physical mass of the
Higgs field: $\Sigma_h(m_h^2)=0$ and $\Sigma_H(m_H^2)=0$
and we solve these equations numerically. 

The wave function renormalization constants $Z_i$, $i=1,2$, have not 
been calculated. We impose that
the residue of the propagator of the CP-even Higgs bosons $h$ and $H$ are 
normalized to one. From eq.~(\ref{eq:eigenvalues}) we get
\begin{eqnarray}
{{\widehat\Sigma_{11}^{\chi}(m_h^2)}\over{Z_2}}+
{{\widehat\Sigma_{22}^{\chi}(m_h^2)}\over{Z_1}}&=&
{{\partial}\over{\partial p^2}}\left\{\widehat\Sigma_{11}^{\chi}(p^2)
\widehat\Sigma_{22}^{\chi}(p^2)-\left[\widehat\Sigma_{11}^{\chi}(p^2)
\right]^2\right\}(m_h^2)\,,
\nonumber\\
{{\widehat\Sigma_{11}^{\chi}(m_H^2)}\over{Z_2}}+
{{\widehat\Sigma_{22}^{\chi}(m_H^2)}\over{Z_1}}&=&
{{\partial}\over{\partial p^2}}\left\{\widehat\Sigma_{11}^{\chi}(p^2)
\widehat\Sigma_{22}^{\chi}(p^2)-\left[\widehat\Sigma_{11}^{\chi}(p^2)
\right]^2\right\}(m_H^2)\,,
\label{eq:hWaveFunct}
\end{eqnarray}
from where the wave function renormalization constants can be calculated.

The matrix ${\bold{\Sigma^{\chi}}}(p^2)$ is diagonalized at a particular
momentum $p^2$ by a rotation defined by the angle $\alpha(p^2)$. 
This mixing angle satisfy
\begin{equation}
\tan2\alpha(p^2)={{2\widehat\Sigma^{\chi}_{12}(p^2)Z_1^{1/2}Z_2^{1/2}
}\over{\widehat\Sigma^{\chi}_{11}(p^2)Z_1-\widehat\Sigma^{\chi}_{22}
(p^2)Z_2}}
\label{eq:tantwoalpha}
\end{equation}
In ref. \cite{yamada} it is used a wave function renormalization of the
type in ref. \cite{michael} to renormalize the CP-even Higgs sector
of the MSSM, obtaining formulas analogous to eq.~(\ref{eq:betas}), and 
then it is defined the angle $\alpha$ at two different scales. Those
two angles are special cases of the momentum dependent mixing
angle $\alpha(p^2)$ in eq.~(\ref{eq:tantwoalpha}).

It is worth to mention that, since the residue of the pole of the 
propagator of the Higgs $A$ is not normalized to unity, when  
renormalizing processes with external Higgs $A$ it is necessary to 
multiply by the finite wave function normalization 
\begin{equation}
Z_A^{1/2}=\left[{\textstyle{{\partial}\over{\partial p^2}}}
\widehat\Sigma^{\varphi}_A(p^2)\right]^{-1/2}_{p^2=m_A^2}
=\left[{{{\textstyle{{\partial}\over{\partial p^2}}}\left(
\widehat\Sigma_{11}^{\varphi}\widehat\Sigma_{22}^{\varphi}-
\widehat\Sigma_{12}^{\varphi 2}\right)Z_1Z_2}\over{
\widehat\Sigma_{11}^{\varphi}Z_1+\widehat\Sigma_{22}^{\varphi}Z_2}}
\right]^{-1/2}_{p^2=m_A^2}
\label{eq:waverenAh}
\end{equation}
which up to one--loop order reduces to $Z_A^{1/2}=1+\half A'_{AA}(m_A^2)$,
as expected.

Finally, if we make the expansion
$Z_i^{1/2}=(1+\delta Z_i)^{1/2}\approx 1+\half\delta Z_i$. We get
\begin{eqnarray}
\Sigma_{11}^{\chi}(p^2)&=&p^2-m_{11}^{\chi 2}+(p^2-m_{11}^{\chi 2})
\delta Z_{H_1}+\delta m_{11}^{\chi 2}-A_{11}^{\chi}(p^2)\nonumber\\
\Sigma_{22}^{\chi}(p^2)&=&p^2-m_{22}^{\chi 2}+(p^2-m_{22}^{\chi 2})
\delta Z_{H_2}+\delta m_{22}^{\chi 2}-A_{22}^{\chi}(p^2)
\label{eq:sigmaelem}\\
\Sigma_{12}^{\chi}(p^2)&=&-m_{12}^{\chi 2}-\half m_{12}^{\chi 2}
(\delta Z_{H_1}+\delta Z_{H_2})+\delta m_{12}^{\chi 2}
-A_{12}^{\chi}(p^2)\nonumber
\end{eqnarray}
And from here we have checked that all the divergences cancel in each
matrix element in eq.~(\ref{eq:sigmaelem}). 

\subsection{Perturbative limit}

At this point, it is instructive to make a perturbative expansion
of $\tan 2\alpha(p^2)$ defined in eq.~(\ref{eq:tantwoalpha}).
If we call $\Delta t_{2\alpha}=\tan2\alpha-t_{2\alpha_0}$, and keeping
only one--loop terms, we get:
\begin{equation}
{{\Delta t_{2\alpha}}\over{t_{2\alpha_0}}}=
-{2\over{(m_A^2+m_Z^2)s_{2\beta}}}\left[\widetilde A_{12}^{\chi}
+\half(\widetilde A_{22}^{\chi}-\widetilde A_{11}^{\chi})t_{2\alpha}
\right]
\label{eq:Dtwoalphapert}
\end{equation}
The variation of the angle $\alpha$ in eq.~(\ref{eq:Dtwoalphapert}) is
the finite one-loop correction to the tree level angle defined
in eq.~(\ref{eq:alphatree}), calculated perturbatively. Now, to clarify
even more the meaning of the angle $\alpha(p^2)$ in 
eq.~(\ref{eq:tantwoalpha}) and the perturbative one-loop correction 
to $\alpha_0$ in
eq.~(\ref{eq:Dtwoalphapert}), consider the inverse propagator matrix
${\bold{\Sigma^{\chi}}}$ with matrix elements given in
eq.~(\ref{eq:sigmaelemnew}). If we rotate ${\bold{\Sigma^{\chi}}}$
by an angle $\alpha_0$ we will diagonalize the tree
level part of the inverse propagator matrix, but not the 
one-loop part:
\begin{equation}
{\bold{R_{\alpha_0}}}{\bold{\Sigma^{\chi}}}(p^2)
{\bold{R^{-1}_{\alpha_0}}}=\left[\matrix{
p^2-m_{H_0}^2-\widetilde A_{H_0H_0}(p^2)&
-\widetilde A_{h_0H_0}(p^2)\cr
-\widetilde A_{h_0H_0}(p^2)&
p^2-m_{h_0}^2-\widetilde A_{h_0h_0}(p^2)\cr}\right]
\label{eq:Sigrotaltree}
\end{equation}
where the rotation matrix is defined by
\begin{equation}
{\bold{R_{\alpha_0}}}=\left[\matrix{
\cos\alpha_0&\sin\alpha_0\cr
-\sin\alpha_0&\cos\alpha_0\cr}\right]\,,
\label{eq:rotmataltree}
\end{equation}
$m_{h_0}^2$ and $m_{H_0}^2$ are the tree level masses given in
eq.~(\ref{eq:Hhtreemass}), $h_0$ and $H_0$ are 
the tree level rotated CP-even fields:
\begin{equation}
{{H_0}\choose{h_0}}={\bold{R_{\alpha_0}}}
{{\chi^0_1}\choose{\chi^0_2}}\,,
\label{eq:rotHhfields}
\end{equation}
and $\widetilde A_{ab}(p^2)$ ($a,b=h_0,H_0$) are the renormalized 
two-point functions. Now, to further diagonalize 
the inverse propagator matrix in eq.~(\ref{eq:Sigrotaltree}) we need a 
rotation of one-loop order by an angle $\Delta\alpha$:
\begin{equation}
{\bold{R_{\Delta\alpha}}}\approx\left[\matrix{
1&\Delta\alpha\cr-\Delta\alpha&1\cr}\right]\,.
\label{eq:rotdeltalpha}
\end{equation}
Imposing that the off-diagonal matrix element in
${\bold{R_{\Delta\alpha}}}{\bold{R_{\alpha_0}}}
{\bold{\Sigma^{\chi}}}(p^2){\bold{R^{-1}_{\alpha_0}}}
{\bold{R^{-1}_{\Delta\alpha}}}$ is zero, we get
\begin{equation}
\Delta\alpha={{c_{2\alpha_0}\widetilde A_{12}+\half s_{2\alpha_0}
(\widetilde A_{22}-\widetilde A_{11})}\over{
m_{H_0}^2-m_{h_0}^2}}
\label{eq:Deltalpha}
\end{equation}
where $c_{2\alpha_0}=\cos2\alpha_0$ and $s_{2\alpha}=\sin2\alpha_0$.
Using the relations $\Delta s_{\alpha}=c_{\alpha_0}\Delta\alpha$,
and $\Delta c_{\alpha}=-s_{\alpha_0}\Delta\alpha$, we can prove
that
\begin{equation}
{{\Delta t_{2\alpha}}\over{t_{2\alpha_0}}}=
{{\Delta\alpha}\over{s_{\alpha_0}c_{\alpha_0}
(c_{\alpha_0}^2-s_{\alpha_0}^2)}}
\label{eq:Dttwoalnew}
\end{equation}
and from eq.~(\ref{eq:Deltalpha}) and using 
$s_{2\alpha_0}(m_{H_0}^2-m_{h_0}^2)=-(m_A^2+m_Z^2)s_{2\beta}$
we recover eq.~(\ref{eq:Dtwoalphapert}). In conclusion, we have proved
that 
\begin{equation}
{\bold{R_{\alpha(p^2)}}}\approx
{\bold{R_{\Delta\alpha}}}{\bold{R_{\alpha_0}}}
\label{eq:Ralphaapprox}
\end{equation}
\ie, a rotation by the angle $\alpha_0$ followed by a rotation
by the angle $\Delta\alpha$ is the perturbative approximation
of a rotation by the angle $\alpha(p^2)$.

\subsection{Relation with the running mixing angle}

Another useful point to clarify here is the relation between
the momentum dependent mixing angle $\alpha(p^2)$ and the 
running mixing angle $\alpha(Q)$, where $Q$ is an arbitrary
mass scale introduced by the momentum subtraction scheme 
$\overline{DR}$ in dimensional reduction \cite{DRED}
\footnote{Or $\overline{MS}$ in the case of 
dimensional regularization used in non-supersymmetric
theories.}. The renormalization group equation (RGE) satisfied
by the angle $\alpha$ is directly related to the divergent
terms of its counterterm. Since $t_{2\alpha_0}$ is defined at
tree level by
\begin{equation}
t_{2\alpha_0}={{2m_{12}^{\chi 2}}\over{m_{11}^{\chi 2}-
m_{22}^{\chi 2}}}\,,
\label{eq:ttwoaltree}
\end{equation}
its counterterm, defined by the relation 
$t_{2\alpha_0}=t_{2\alpha}-\delta t_{2\alpha}$, satisfy
\begin{equation}
{{\delta t_{2\alpha}}\over{t_{2\alpha}}}=-{2\over{
(m_A^2+m_Z^2)s_{2\beta}}}\left[\delta m_{12}^{\chi 2}
+\half(\delta m_{22}^{\chi 2}-\delta m_{11}^{\chi 2})
t_{2\alpha}\right]\,.
\label{eq:countttwoal}
\end{equation}
This term is contained in the finite shift of the angle $\alpha$
given in eq.~(\ref{eq:Dtwoalphapert}), then we have
\begin{eqnarray}
{{\Delta t_{2\alpha}}\over{t_{2\alpha}}}=-{{\delta t_{2\alpha}}
\over{t_{2\alpha}}}-{2\over{(m_A^2+m_Z^2)s_{2\beta}}}\Bigg[&&
\half(m_{22}^{\chi 2}\delta Z_{H_2}-m_{11}^{\chi 2}\delta Z_{H_1})
t_{2\alpha}+\nonumber\\ &&
\half m_{12}^{\chi 2}(\delta Z_{H_1}+\delta Z_{H_2})+
{{A_{h_0H_0}(0)}\over{c_{2\alpha}}}\Bigg]\,.
\label{eq:deltcounttrel}
\end{eqnarray}
Since $\Delta t_{2\alpha}$ is finite, to find the divergent terms
of the counterterm $\delta t_{2\alpha}$ it is enough to look for 
the divergences of the second term in the right hand side of 
eq.~(\ref{eq:deltcounttrel}). In order to compare with ref.~\cite{pomarol},
where the contribution from the top quark to the counterterm 
for the angle $\alpha$ is calculated, we concentrate only in the 
top quark contribution. From eq.~(\ref{eq:hWaveFunct}), and keeping
only one--loop divergent terms, we have
\begin{equation}
\Big[\delta Z_{H_1}\Big]^{top}_{div}=0\,,\qquad
\Big[\delta Z_{H_2}\Big]^{top}_{div}=
-{{N_cg^2m_t^2}\over{32\pi^2m_W^2s_{\beta}^2}}\Delta\,,
\label{eq:DeltasZtop}
\end{equation}
where $N_c=3$ is the number of colors, $\Delta$ is the regulator 
in dimensional regularization given by
\begin{equation}
\Delta={2\over{4-n}}+\ln 4\pi-\gamma_E,
\label{eq:Deltareg}
\end{equation}
$n$ is the number of space-time dimensions, and $\gamma_E$ is
the Euler's constant. On the other hand, the contribution from
the top quark to the mixing between $h_0$ and $H_0$ is
\begin{equation}
\Big[A_{h_0H_0}(0)\Big]^{top}_{div}=
{{3N_cg^2s_{2\alpha}m_t^4}\over{32\pi^2m_W^2s_{\beta}^2}}
\Delta\,.
\label{eq:hHdiv}
\end{equation}
Replacing eqs.~(\ref{eq:DeltasZtop}) and (\ref{eq:hHdiv}) into 
eq.~(\ref{eq:deltcounttrel}),
and considering that $\Delta t_{2\alpha}$ is finite, we get
\begin{equation}
\left[{{\delta t_{2\alpha}}\over{t_{2\alpha}}}
\right]^{top}_{div}=-{{N_cg^2m_t^2}\over{64\pi^2m_W^2s_{\beta}^2}}
{{12m_t^2-m_A^2-m_Z^2}\over{(m_A^2+m_Z^2)s_{2\beta}}}t_{2\alpha}
\Delta
\label{eq:deltatdiv}
\end{equation}
and this is the same answer we find in ref.~\cite{pomarol}, with
the exception of the sign, since here we define the counterterm
of the angle $\alpha$ with the opposite sign.

Now we have checked that we reproduce the correct divergent terms for
the $\tan2\alpha$ counterterm, we turn to the relation itself
between the renormalized $\tan2\alpha(p^2)$ and the running
$\tan2\alpha(Q)$. In the $\overline{MS}$ scheme, each counterterm
is fixed to cancel the divergent pieces of the corresponding
loop corrections. In the case of tadpoles we have
\begin{equation}
\delta t_i^{\overline{MS}}=[T_i(Q)]^{div}
\qquad {\mathrm{and}} \qquad
-\delta t_i^{\overline{MS}}+T_i(Q)\equiv\widetilde{T}_i^
{\overline{MS}}(Q)\,.
\label{eq:RenTad}
\end{equation}
Therefore, the running tadpoles are equal to\footnote{
To see the effect the running tadpoles have in the determination
of running parameters, see \cite{marcotadpoles}.
}
\begin{eqnarray}
t_1(Q)&\equiv&\left[m_{1H}^2v_{1}-m_{12}^2v_{2}+
\eighth(g^2+g'^2)v_{1}(v_{1}^2-v_{2}^2)\right](Q)=
-\widetilde{T}_1^{\overline{MS}}(Q)
\,,\nonumber\\
t_2(Q)&\equiv&\left[m_{2H}^2v_{2}-m_{12}^2v_{1}+
\eighth(g^2+g'^2)v_{2}(v_{2}^2-v_{1}^2)\right](Q)=
-\widetilde{T}_2^{\overline{MS}}(Q)\,.
\label{eq:RunTad}
\end{eqnarray}
Considering the tree level neutral CP--even Higgs mass
matrix in eq.~(\ref{eq:cpevmasmat}), we find that the renormalized
inverse propagator matrix in the $\overline{MS}$ scheme has
the following matrix elements:
\begin{eqnarray}
\Sigma_{11}^{\chi}(p^2)&=&p^2-m_Z^2(Q)c_{\beta}^2(Q)-m_A^2(Q)
s_{\beta}^2(Q)+{1\over{v_1}}\widetilde{T}_1^{\overline{MS}}(Q)
-\widetilde A_{11}^{\chi\overline{MS}}(p^2,Q)\nonumber\\
\Sigma_{22}^{\chi}(p^2)&=&p^2-m_Z^2(Q)s_{\beta}^2(Q)-m_A^2(Q)
c_{\beta}^2(Q)+{1\over{v_2}}\widetilde{T}_2^{\overline{MS}}(Q)
-\widetilde A_{22}^{\chi\overline{MS}}(p^2,Q)
\label{eq:sigmaelemMS}\\
\Sigma_{12}^{\chi}(p^2)&=&[m_Z^2(Q)+m_A^2(Q)]s_{\beta}(Q)c_{\beta}(Q)-
\widetilde A_{12}^{\chi\overline{MS}}(p^2,Q)\nonumber
\end{eqnarray}
and from here we can find the renormalized $\alpha(p^2)$ in terms of
$\overline{MS}$ quatities
\begin{eqnarray}
&&\tan2\alpha(p^2)=\label{eq:tantwoalphaOSMS}\\
&&{{[m_A^2(Q)+m_Z^2(Q)]s_{2\beta}(Q)-
2\widetilde A_{12}^{\chi\overline{MS}}(p^2,Q)
}\over{[m_A^2(Q)-m_Z^2(Q)]c_{2\beta}(Q)-
\widetilde A_{11}^{\chi\overline{MS}}(p^2,Q)
+\widetilde A_{22}^{\chi\overline{MS}}(p^2,Q)
+{1\over{v_1}}\widetilde{T}_1^{\overline{MS}}(Q)
-{1\over{v_2}}\widetilde{T}_2^{\overline{MS}}(Q)}}\nonumber
\end{eqnarray}
or, in first approximation
\begin{eqnarray}
\tan2\alpha(p^2)&\approx &t_{2\alpha}(Q)-{2\over{(m_A^2-m_Z^2)c_{2\beta}}}
\Bigg\{\widetilde{A}^{\chi\overline{MS}}_{12}(p^2,Q)
\label{eq:tantwoalappOSMS}\\&+&{1\over2}\left[
\widetilde{A}^{\chi\overline{MS}}_{22}(p^2,Q)-\widetilde{A}^{\chi
\overline{MS}}_{11}(p^2,Q)-{1\over{v_2}}\widetilde{T}_2^{\overline{MS}}(Q)
+{1\over{v_1}}\widetilde{T}_1^{\overline{MS}}(Q)\right]t_{2\alpha}\Bigg\}
\nonumber
\end{eqnarray}
where the running $\tan2\alpha$ is defined as
\begin{equation}
t_{2\alpha}(Q)={{m_A^2(Q)+m_Z^2(Q)}\over{m_A^2(Q)-m_Z^2(Q)}}
t_{2\beta}(Q)\,.
\label{eq:tantwoalphaMS}
\end{equation}
The relation between $\tan2\alpha(p^2)$ and $\tan2\alpha(Q)$ in
eq.~(\ref{eq:tantwoalphaOSMS}) is completed when we give the on-shell
definition of the $Z$ and $A$ masses:
\begin{equation}
m_Z^2=m_Z^2(Q)+{\rm Re}\widetilde{A}_{ZZ}^{\overline{MS}}(m_Z^2,Q),
\qquad
m_A^2=m_A^2(Q)+{\rm Re}\widetilde{A}_{AA}^{\overline{MS}}(m_A^2,Q),
\label{eq:polemZmA}
\end{equation}
where the tilde over the self energies means that the counterterms
have already canceled the divergent terms. The relation between the 
on-shell definition of $t_{\beta}$ and the running $t_{\beta}(Q)$
can be deduced from ref.~\cite{diazHdecay}.

\subsection{Numerical Results}

In this section we compare numerically the momentum dependent mixing angle
method of diagonalizing coupled scalars with more conventional methods.
We have chosen to compare with:

\noindent (a) the tree level approximation;

\noindent (b) the approximation where
we diagonalize the matrix formed by the
tree level quantities plus the pieces of the radiative corrections
proportional to $m_t^4$ \cite{BarbieriF,diazhaberii}. By doing this we 
obtain the eigenvalues
\begin{equation}
\widehat{m}^2_{H,h}={\textstyle{1\over 2}}(m_A^2+m_Z^2+\Delta_t)\pm
{\textstyle{1\over 2}}\sqrt{[(m_Z^2-m_A^2)c_{2\beta}-\Delta_t]^2
+(m_A^2+m_Z^2)^2s_{2\beta}^2}\,,
\label{eq:exactdiag}
\end{equation}
and mixing angle
\begin{equation}
\tan 2\widehat{\alpha}={{(m_A^2+m_Z^2)s_{2\beta}}\over{
(m_A^2-m_Z^2)c_{2\beta}+\Delta_t}}\,.
\label{eq:alphabar}
\end{equation}
where:
\begin{equation}
\Delta_t={{3g^2m_t^4}\over{16\pi^2m_W^2s_{\beta}^2}}\,\ln\,
{{m_{\tilde t_1}^2m_{\tilde t_2}^2}\over{m_t^4}}\,;
\label{eq:deltat}
\end{equation}
and

\noindent (c) the leading logarithms approximation.

The exact formulae for loops involving top and bottom quarks
and squarks are given in the appendix (see also ref.~\cite{thesis}), 
and loops corresponding to
the gauge bosons, Higgs bosons, charginos and neutralinos are treated
at the leading logarithm approximation.
It is useful to evaluate the formulae in the appendix in the limit
$M_{SUSY}^2\gg m_t^2\gg m_Z^2\gsim m_A^2\gg m_b^2$ and $\mu=A_U=A_D=0$,
where all the squark soft supersymmetry breaking mass parameters are
assumed to be of the order of $M_{SUSY}$. The leading logarithms
obtained in this way are in agreement with ref.~\cite{rgegeneral}:
\begin{eqnarray}
\Sigma_{11}^{\chi}(p^2)&=&p^2-m_Z^2c_{\beta}^2-m_A^2s_{\beta}^2
-{{3g^2}\over{16\pi^2m_W^2}}\left({{2m_b^4}\over{c_{\beta}^2}}
-m_Z^2m_b^2\right)\ln{{M_{SUSY}^2}\over{m_{weak}^2}}\nonumber\\
&-&{{g^2m_Z^2c_{\beta}^2}\over{32\pi^2c_W^2}}(P_t+P_b)\ln
{{M_{SUSY}^2}\over{m_{weak}^2}}+{\cal O}(g^2m_Z^2)\nonumber\\
\nonumber\\
\Sigma_{22}^{\chi}(p^2)&=&p^2-m_Z^2s_{\beta}^2-m_A^2c_{\beta}^2
-{{3g^2}\over{16\pi^2m_W^2}}\left({{2m_t^4}\over{s_{\beta}^2}}
-m_Z^2m_t^2\right)\ln{{M_{SUSY}^2}\over{m_{weak}^2}}\nonumber\\
&-&{{g^2m_Z^2s_{\beta}^2}\over{32\pi^2c_W^2}}(P_t+P_b)\ln
{{M_{SUSY}^2}\over{m_{weak}^2}}-{{g^2m_t^2}\over{16\pi^2m_W^2
s_{\beta}^2}}p^2+{\cal O}(g^2m_Z^2)
\label{eq:sabappr}\\ \nonumber\\
\Sigma_{12}^{\chi}(p^2)&=&s_{\beta}c_{\beta}(m_Z^2+m_A^2)
-{{3g^2}\over{32\pi^2c_W^2}}\left({{m_t^2}\over{t_{\beta}}}
+m_b^2t_{\beta}\right)\ln{{M_{SUSY}^2}\over{m_{weak}^2}}\nonumber\\
&+&{{g^2m_Z^2s_{\beta}c_{\beta}}\over{32\pi^2c_W^2}}(P_t+P_b)\ln
{{M_{SUSY}^2}\over{m_{weak}^2}}+{\cal O}(g^2m_Z^2)\nonumber
\end{eqnarray}
where $P_t=1-4e_ts_W^2+8e_t^2s_W^4$ and $P_b=1+4e_bs_W^2+8e_b^2
s_W^4$. Note that it also displayed the largest of the non-leading
logarithm terms; it is proportional to the momentum $p^2$ and
to the second power of the top quark mass. 

The $hZZ$ coupling in the MSSM relative to the same coupling in the
SM is given by the parameter $\sin(\beta-\alpha)$. In Fig.~1 we plot 
this parameter as a function of $\tan\beta$. The upper dotdashed curve
correspond to the tree level approximation. The $\Delta_t$--improved
approximation is in the dotted line [calculated with $\widehat\alpha$ 
defined in eq.~(\ref{eq:alphabar})]. The leading logarithms approximation
is plotted in the lower dotdashed line. The parameter $\sin(\beta-\alpha)$
calculated with the momentum dependent mixing angle $\alpha(p^2)$ is
plotted in the dashed curves for $p^2=m_H^2$ and in the solid curves for
$p^2=m_h^2$, for the cases: (a) $A=\mu=1$ TeV and (b) $A=-\mu=1$ TeV,
which define the squark mixing.

{}From Fig.~1 we can learn that the tree level approximation can give
completely wrong results. The $\Delta_t$--improved and the leading 
logarithms approximations give results that are quite close to each 
other, indicating that the $m_t^4$ terms are the main contributions 
to the leading logarithms for the chosen parameters of this figure. 
Both approximations can differ from the results using $\alpha(p^2)$, 
specially in the high $\tan\beta$ region. The parameter calculated 
with our method shows a strong dependence on the squark mixing 
parameters, as it can be appreciated from the solid and dashed curves 
in cases (a) and (b). We also can see that the angle $\alpha(p^2)$ 
has a small variation between the two physical external momenta 
$p^2=m_h^2$ and $p^2=m_H^2$.

The same kind of graph is in Fig.~2, where we plot $\sin(\beta-\alpha)$
as a function of $\tan\beta$. As opposed to the previous figure, here
we consider lighter squarks: $M_Q=M_U=M_D\equiv M_{SUSY}=200$ GeV, 
$A_U=A_D\equiv A=140$ GeV, and $\mu=-70$ GeV. The most interesting 
feature here is that the parameter $\sin(\beta-\alpha)$ is substantially
different in the two different external momenta $p^2=m_h^2$ and 
$p^2=m_H^2$ when $\tan\beta$ is large. In addition, this time not only
the tree level calculation gives wrong results, but also the leading
logarithms approximation. The fact that squarks are light implies that
leading logarithms of particles other than squarks are also important,
and this is reflected in the fact that the $\Delta_t$--improved
approximation is very different from the complete leading logarithm
approximation. For the parameters chosen in this figure, we see that the 
$\Delta_t$--improved curve is close to the curves calculated with 
$\alpha(p^2)$, but this is an accident as we can see in the next figure.

In Fig.~3 we plot $\sin(\beta-\alpha)$ as a function of $m_A$ for a fixed
value of $\tan\beta=40$ and for light squarks. Here it become evident that
any of the three traditional approximations can give a value of
$\sin(\beta-\alpha)$ with an error of 40\% or more. Considering that the 
relevant parameter for the Higgs search is the MSSM is 
$\sin^2(\beta-\alpha)$ we see that the error on the cross section
can be of the order of 60\% !

The Higgs mass calculated with our method was compared to more traditional
methods in ref.~\cite{Vanderbilt}. In that reference we mention the
differences between the Effective Potential method, the Renormalization
Group Equations method, and the Diagramatic method. Here we discuss
quantitatively these differences through an example given by the choice
of parameters in Fig.~4. 

The Higgs masses are calculated by finding the zeros of the determinant of
the inverse propagator matrix ${\bold{\Sigma^{\chi}}}(p^2)$ as indicated by 
eq.~(\ref{eq:detsigmazero}). The non--trivial momentum $p^2$
dependence of the matrix elements $\widehat\Sigma_{ij}^{\chi}(p^2)$
is in the self energies $A_{ij}^{\chi}(p^2)$. Usually, the momentum squared
in the self energies is replaced by a constant. For example, in the
effective potential method the momentum squared in the self energies
is replaced by zero. Another typical choice is to take $p^2$ equal to the
tree level mass. In Fig.~4 we have replaced $p^2$ by the same constant
$\overline{p}^2$ in the three self energies and plot as a function of this 
constant the two Higgs masses $m_h$ and $m_H$ calculated after that 
replacement. In the solid (dashed) line we have $m_h$ ($m_H$) as a function 
of the squared root of the argument of the self energies.

We see that, for the choice of parameters in Fig.~4, the dependence of
the Higgs masses $m_h$ and $m_H$ on $\overline{p}^2$ is quite strong,
specially for large values of $\overline{p}^2$. The dotted line is the
diagonal defined by $m=\sqrt{\overline{p}^2}$. The intersection of this line 
with the solid and the dashed curves give us the pole masses $m_h$ and $m_H$
calculated with our method. These values are $m_h=64.8$ GeV and $m_H=95.4$
GeV. On the other hand, the intersection of the solid and dashed curves
with the vertical line defined by $\overline{p}=0$ corresponds to the
masses calculated with the second derivative of the effective potential. 
These values are $m_h^{eff}=60.8$ GeV and $m_H^{eff}=108.4$ GeV, and they
are different from the pole masses. Therefore, it is clear that if we replace 
the momentum $p^2$ in the self energies by a constant, the Higgs masses
calculated in that way may depend strongly on that choice. To be 
complete, we give the value of the Higgs masses calculated in 
(a) the tree level approximation $m_h^{tree}=90.7$ GeV and 
$m_H^{tree}=100.4$ GeV, (b) the $\Delta_t$--improved approximation
$m_h^{\Delta_t}=99.9$ GeV and $m_H^{\Delta_t}=129.3$ GeV, and (c) the
leading logarithms approximation $m_h^{l.l.}=100.0$ GeV and 
$m_H^{l.l.}=133.9$ GeV, valid for the choice of parameters in Fig.~4.
The reason why the Higgs masses calculated with these approximations
differs so much from the masses calculated with our method is the
large value of the squark mixing: approximations (a) to (c) do not
treat appropriately the squark mixing. An approximated formula which
treats the squark mixing can be found for example in ref.~\cite{CEQWyHHH}.
As it can be seen above, a better
value is obtained with the effective potential, but still, differences can
be of 6\% to 14\%. It could be argued that the effective potential 
method can be improved by correcting the fact that it is found at zero
external momentum, \ie, corrections of the type 
$\Delta m^2=A_{hh}(m^2_{h,tree})-A_{hh}(0)$. Nevertheless, this correction
implies that all the information comming from the effective potential
is canceled out, and we are left with the pure diagramatic method. Therefore,
we could have started with the diagramatic method in the first place
and forget about the effective potential. The necessity of the above 
correction disappears in first approximation if $m_h^{tree}=0$, as it 
was done in ref.~\cite{diazhaberii} (see also discussion in 
ref.~\cite{Vanderbilt}).

\section{Conclusions}

We have developed a new method of diagonalizing two coupled scalars by 
introducing a momentum dependent mixing angle $\alpha(p^2)$, where $p$
is the external momentum of the two--point functions. The dependence of the 
mixing angle on the momentum $p^2$ indicate us that the rotation matrix
which diagonalizes the inverse propagator matrix is different whether we
are at the pole mass $p^2=m_h^2$ or $p^2=m_H^2$ or at any other scale. 
We have compared this method with the conventional wave function 
renormalization and with the mixed wave function renormalization In fact, we 
introduced the momentum dependent mixing angle as a generalization of the 
previous methods.

We applied this method to the diagonalization of the CP--even Higgs
bosons inverse propagator matrix in the MSSM. 
We used the diagramatic method in 
an on--shell renormalization scheme, where the tadpoles are exactly
zero at one--loop, the masses $m_Z$, $m_W$, and $m_A$ are defined as
the pole masses, and $\tan\beta$ is defined through the on--shell
definition of the slepton masses $m_{\tilde e_L}$ and $m_{\tilde\nu_e}$.
We calculate the wave function renormalization constants $Z_1$ and $Z_2$
by imposing that the residue of the propagators of the CP--even Higgs
bosons are exactly one. We do this by using a formula for $Z_1$ and $Z_2$
valid for any number of loops and, therefore, specially useful when 
radiative corrections are large. We make explicit the relation between 
the momentum dependent mixing angle, which includes some effects of
higher order loops, to the mixing angle calculated in the exact one--loop
perturbative limit. We also make explicit the relation between the 
momentum dependent mixing angle $\alpha(p^2)$ and the running mixing
angle $\alpha(Q)$, where $Q$ is the arbitrary mass scale of the 
$\overline{MS}$ scheme. We give some numerical results by calculating 
the parameter $\sin(\beta-\alpha)$, which is the ratio between the 
$ZZh$ coupling in the MSSM to the same coupling in the SM. We compare this
parameter calculated with the momentum dependent mixing angle 
$\alpha(p^2)$ with (a) the tree level approximation, (b) the 
$\Delta_t$--improved approximation (including only terms proportional
to $m_t^4$), and (c) the leading logarithms approximation. We find 
important numerical differences between the different methods, and
they are relevant for the Higgs searches at LEP2 in the region of
parameter space where $m_A={\cal O}(m_Z)$.

Finally, we calculate the pole masses of the CP-even Higgs bosons
$m_h$ and $m_H$. We show that they are exactly independent of the
wave function renormalization constants $Z_1$ and $Z_2$ only if we
use the formulas in eq.~(\ref{eq:hWaveFunct}). The Higgs masses are 
calculated by finding the zeros of the determinant of the inverse 
propagator matrix [eq.~(\ref{eq:detsigmazero})], and this is the direct 
consequence of the definition of the momentum dependent mixing angle 
$\alpha(p^2)$. We compare with the Higgs masses calculated in the three 
approximations described in the above paragraph and we find that for some 
choices of parameter space there are non--negligible differences 
between them, and therefore, in those cases our method should be used.

It is worth to mention also that our method can be easily generalized to
the diagonalization of more than two coupled scalars. Also the 
generalization to the diagonalization of coupled fermions or coupled 
vector bosons is straight forward.

\section*{Acknowledgements}

I am thankful to H.E. Haber for his help in the early stages
of this work. Useful conversations with R. Hempfling, 
S. King, and T. Weiler are specially appreciated.
Part of this work was done at Vanderbilt University, 
part at the University of Southampton, and part at the
ICTP. I was supported in Spain by a DGICYT postdoctoral grant.

\section*{Appendix}

In this appendix we display the exact one-loop formulae we need to
compute the inverse propagator matrix for the CP-even Higgs fields,
for loops involving top and bottom quarks and squarks.
We exhibit separately the contribution from the quarks and 
from the squarks loops.

To obtain the mass counterterms $\delta m_{ij}^2$ given in
eq.~(\ref{eq:masscountchi}), the mass counterterm of the $Z$ gauge boson
is required. It is given by $\delta m_Z^2=A_{ZZ}(m_Z^2)$
where the contribution from top and bottom quarks is:
\begin{eqnarray}
\Big[A_{ZZ}(m_Z^2)\Big]^{tb}&=&
{{N_cg^2}\over{32\pi^2c_W^2}}(m_t^2B_0^{Ztt}+m_b^2B_0^{Zbb})
\nonumber\\
&-&{{N_cg^2}\over{16\pi^2c_W^2}}(\quarter-e_ts_W^2+2e_t^2s_W^4)
(4B_{22}^{Ztt}-2A_0^t+m_Z^2B_0^{Ztt})
\nonumber\\
&-&{{N_cg^2}\over{16\pi^2c_W^2}}(\quarter+e_bs_W^2+2e_b^2s_W^4)
(4B_{22}^{Zbb}-2A_0^b+m_Z^2B_0^{Zbb})
\label{eq:ZZtb}
\end{eqnarray}
where we use the notation $B_{22}^{Ztt}\equiv B_{22}(m_Z^2;m_t^2,m_t^2)$, 
and similarly for the other Veltman's functions. The contribution to
the $Z$ self energy from top and bottom squarks is:
\begin{eqnarray}
\Big[A_{ZZ}(m_Z^2)\Big]^{\tilde t\tilde b}&=&
{{N_cg^2}\over{8\pi^2c_W^2}}\Big[
\nonumber\\
&&(-\half c_t^2+e_ts_W^2)^2(2B_{22}^{Z\tilde t_1\tilde t_1}
-A_0^{\tilde t_1})+
(\half c_b^2+e_bs_W^2)^2(2B_{22}^{Z\tilde b_1\tilde b_1}
-A_0^{\tilde b_1})
\nonumber\\
&+&(-\half s_t^2+e_ts_W^2)^2(2B_{22}^{Z\tilde t_2\tilde t_2}
-A_0^{\tilde t_2})+
(\half s_b^2+e_bs_W^2)^2(2B_{22}^{Z\tilde b_2\tilde b_2}
-A_0^{\tilde b_2})
\nonumber\\
&+&\quarter s_t^2c_t^2(4B_{22}^{Z\tilde t_1
\tilde t_2}-A_0^{\tilde t_1}-A_0^{\tilde t_2})+
\quarter s_b^2c_b^2(4B_{22}^{Z\tilde b_1
\tilde b_2}-A_0^{\tilde b_1}-A_0^{\tilde b_2})\Big].
\label{eq:ZZstb}
\end{eqnarray}
Here we use the notation $s_t\equiv\sin\alpha_t$ and $c_t\equiv\cos\alpha_t$
where $\alpha_t$ is the rotation angle necessary to diagonalized the top
squark mass matrix. Similar expressions are used for the sbottom mixing
angle $\alpha_b$.

The $W$--boson self energy is
\begin{equation}
\Big[A_{WW}(m_W^2)\Big]^{tb}=
-{{N_cg^2}\over{32\pi^2}}\left[4B_{22}^{Wtb}-A_0^t-A_0^b+
(m_W^2-m_t^2-m_b^2)B_0^{Wtb}\right]
\label{eq:WWtb}
\end{equation}
for the quarks contribution, and 
\begin{eqnarray}
\Big[A_{WW}(m_W^2)\Big]^{\tilde t\tilde b}&=&
{{N_cg^2}\over{32\pi^2}}\Big[
c_t^2c_b^2(4B_{22}^{W\tilde t_1\tilde b_1}-A_0^{\tilde t_1}-A_0^{\tilde b_1})+
c_t^2s_b^2(4B_{22}^{W\tilde t_1\tilde b_2}-A_0^{\tilde t_1}-A_0^{\tilde b_2})
\nonumber\\&+&
s_t^2c_b^2(4B_{22}^{W\tilde t_2\tilde b_1}-A_0^{\tilde t_2}-A_0^{\tilde b_1})+
s_t^2s_b^2(4B_{22}^{W\tilde t_2\tilde b_2}-A_0^{\tilde t_2}-A_0^{\tilde b_2})
\Big]
\label{eq:WWstsb}
\end{eqnarray}
for the squark contributions.
xxx
The third mass counterterm we need to compute
is $\delta(m_{12}^2/s_{\beta}c_{\beta})$ and it is
related to the CP-odd self energy and the tadpoles through
eqs.~(\ref{eq:mcountcpodd}) and (\ref{eq:Arencond}). Thus we need the 
following quantity:
\begin{equation}
\bigg[A_{AA}(m_A^2)-s_{\beta}^2{{\delta t_1}\over{v_1}}
-c_{\beta}^2{{\delta t_2}\over{v_2}}\bigg]^{tb}=
-{{N_cg^2m_A^2}\over{32\pi^2m_W^2}}\Big({{m_t^2}\over{t_{\beta}^2
}}B_0^{Att}+m_b^2t_{\beta}^2B_0^{Abb}\Big)
\label{eq:AAtadtb}
\end{equation}
for top and bottom quark loops, and:
\begin{eqnarray}
&&\bigg[A_{AA}(m_A^2)-s_{\beta}^2{{\delta t_1}\over{v_1}}
-c_{\beta}^2{{\delta t_2}\over{v_2}}\bigg]^{\tilde t\tilde b}=
-{{N_cg^2}\over{32\pi^2m_W^2}}\Big[
\nonumber\\
&&\qquad\qquad
m_t^2(\mu+A_U/t_{\beta})^2B_0^{A\tilde t_1\tilde t_2}
+m_t(\mu t_{\beta}-A_U/t_{\beta}^2)
s_tc_t(A_0^{\tilde t_1}-A_0^{\tilde t_2})
\nonumber\\
&&\qquad\,\,\,\,
+\,\,\,m_b^2(\mu+A_Dt_{\beta})^2B_0^{A\tilde b_1\tilde b_2}
+m_b(\mu/t_{\beta}-A_Dt_{\beta}^2)
s_bc_b(A_0^{\tilde b_1}-A_0^{\tilde b_2})\Big]
\label{eq:AAtadstb}
\end{eqnarray}
for top and bottom squarks loops.

Finally, we need the two point functions $A_{ij}^{\chi}$.
In the case of the self energies, we add the appropriate
tadpole as indicated bellow. 
The contribution due to the top and bottom quarks is very simple:
\begin{eqnarray}
\Big[A_{11}^{\chi}(p^2)-{{\delta t_1}\over{v_1}}\Big]^{tb}&=&
{{N_cg^2m_b^2}\over{32\pi^2m_W^2c_{\beta}^2}}(4m_b^2-p^2)B_0^{pbb}
\nonumber\\
\Big[A_{22}^{\chi}(p^2)-{{\delta t_2}\over{v_2}}\Big]^{tb}&=&
{{N_cg^2m_t^2}\over{32\pi^2m_W^2s_{\beta}^2}}(4m_t^2-p^2)B_0^{ptt}
\nonumber\\
\Big[A_{12}^{\chi}(p^2)\Big]^{tb}&=& 0\,.
\label{eq:chiabtb}
\end{eqnarray}
where we notice that $\chi_1$ couples only to top quarks, and $\chi_2$
couples only to bottom quarks, implying that the mixing is zero.

The contribution due to top and bottom squarks to the $\chi_1\chi_1$
self energy minus the $\chi_1$ tadpole is:
\begin{eqnarray}
\Bigg[A_{11}^{\chi}(p^2)\!\!\!&-&\!\!\!
{{\delta t_1}\over{v_1}}\Bigg]^{
\tilde t\tilde b}=-{{N_cg^2}\over{16\pi^2m_W^2}}\Bigg\{
\left[m_Z^2c_{\beta}(-\half c_t^2+e_ts_W^2c_{2t})+{{m_t
\mu}\over{2s_{\beta}}}s_{2t}\right]^2B_0^{p\tilde t_1
\tilde t_1}
\nonumber\\
&+&\left[m_Z^2c_{\beta}(\half s_t^2+e_ts_W^2c_{2t})+{{m_t
\mu}\over{2s_{\beta}}}s_{2t}\right]^2B_0^{p\tilde t_2
\tilde t_2}
\nonumber\\
&+&2\left[m_Z^2c_{\beta}(\quarter-e_ts_W^2)s_{2t}+{{m_t
\mu}\over{2s_{\beta}}}c_{2t}\right]^2B_0^{p\tilde t_1
\tilde t_2}
+{{m_t\mu}\over{4s_{\beta}c_{\beta}}}s_{2t}(A_0^{\tilde t_1}
-A_0^{\tilde t_2})
\nonumber\\
&+&\left[m_Z^2c_{\beta}(\half c_b^2+e_bs_W^2c_{2b})-{{m_b
A_D}\over{2c_{\beta}}}s_{2b}-{{m_b^2}\over{c_{\beta}}}
\right]^2B_0^{p\tilde b_1\tilde b_1}
\label{eq:chiaastb}\\
&+&\left[m_Z^2c_{\beta}(\half s_b^2-e_bs_W^2c_{2b})+{{m_b
A_D}\over{2c_{\beta}}}s_{2b}-{{m_b^2}\over{c_{\beta}}}
\right]^2B_0^{p\tilde b_2\tilde b_2}
\nonumber\\
&+&2\left[m_Z^2c_{\beta}(\quarter+e_bs_W^2)s_{2b}+{{m_b
A_D}\over{2c_{\beta}}}c_{2b}\right]^2B_0^{p\tilde b_1
\tilde b_2}
-{{m_bA_D}\over{4c_{\beta}^2}}s_{2b}(A_0^
{\tilde b_1}-A_0^{\tilde b_2})\Bigg\},
\nonumber
\end{eqnarray}
where we use the notation $s_{2t}\equiv\sin(2\alpha_t)$, 
$c_{2t}\equiv\cos(2\alpha_t)$, and similarly for $\alpha_b$.
Similarly, the contribution to the $\chi_2\chi_2$ self energy minus 
the $\chi_2$ tadpole, due to top and bottom squarks is:
\begin{eqnarray}
\Bigg[A_{22}^{\chi}(p^2)\!\!\!&-&\!\!\!
{{\delta t_2}\over{v_2}}\Bigg]^{\tilde t\tilde b}=
\nonumber\\
&-&{{N_cg^2}\over{16\pi^2m_W^2}}\Bigg\{
\left[m_Z^2s_{\beta}(-\half c_t^2+e_ts_W^2c_{2t})+{{m_t
A_U}\over{2s_{\beta}}}s_{2t}+{{m_t^2}\over{s_{\beta}}}
\right]^2B_0^{p\tilde t_1\tilde t_1}
\nonumber\\
&+&\left[m_Z^2s_{\beta}(\half s_t^2+e_ts_W^2c_{2t})+{{m_t
A_U}\over{2s_{\beta}}}s_{2t}-{{m_t^2}\over{s_{\beta}}}
\right]^2B_0^{p\tilde t_2\tilde t_2}
\nonumber\\
&+&2\left[m_Z^2s_{\beta}(-\quarter+e_ts_W^2)s_{2t}-{{m_t
A_U}\over{2s_{\beta}}}c_{2t}\right]^2B_0^{p\tilde t_1
\tilde t_2}
-{{m_tA_U}\over{4s_{\beta}^2}}s_{2t}(A_0^{\tilde t_1}
-A_0^{\tilde t_2})
\nonumber\\
&+&\left[m_Z^2s_{\beta}(\half c_b^2+e_bs_W^2c_{2b})-{{m_b
\mu}\over{2c_{\beta}}}s_{2b}
\right]^2B_0^{p\tilde b_1\tilde b_1}
\label{eq:chibbstb}\\
&+&\left[m_Z^2s_{\beta}(-\half s_b^2+e_bs_W^2c_{2b})-{{m_b
\mu}\over{2c_{\beta}}}s_{2b}
\right]^2B_0^{p\tilde b_2\tilde b_2}
\nonumber\\
&+&2\left[m_Z^2s_{\beta}(\quarter+e_bs_W^2)s_{2b}+{{m_b
\mu}\over{2c_{\beta}}}c_{2b}\right]^2B_0^{p\tilde b_1
\tilde b_2}
+{{m_b\mu}\over{4s_{\beta}c_{\beta}}}s_{2b}(A_0^
{\tilde b_1}-A_0^{\tilde b_2})\Bigg\}
\nonumber
\end{eqnarray}
Finally, the $\chi_1\chi_2$ mixing due to squarks is given by:
\begin{eqnarray}
&&\bigg[A_{12}^{\chi}(p^2)\bigg]^{\tilde t\tilde b}=-
{{N_cg^2}\over{16\pi^2m_W^2}}\Bigg\{
-\left[m_Z^2c_{\beta}(-\half c_t^2+e_ts_W^2c_{2t})+{{m_t
\mu}\over{2s_{\beta}}}s_{2t}\right]
\nonumber\\
&&\qquad\qquad\qquad\qquad\qquad\times
\left[m_Z^2s_{\beta}(-\half c_t^2+e_ts_W^2c_{2t})+{{m_t
A_U}\over{2s_{\beta}}}s_{2t}+{{m_t^2}\over{s_{\beta}}}
\right]B_0^{p\tilde t_1\tilde t_1}
\nonumber\\
&+&2\left[m_Z^2c_{\beta}(\quarter-e_ts_W^2)s_{2t}+{{m_t
\mu}\over{2s_{\beta}}}c_{2t}\right]
\left[m_Z^2s_{\beta}(-\quarter+e_ts_W^2)s_{2t}-{{m_t
A_U}\over{2s_{\beta}}}c_{2t}\right]B_0^{p\tilde t_1
\tilde t_2}
\nonumber\\
&-&\left[m_Z^2c_{\beta}(\half s_t^2+e_ts_W^2c_{2t})+{{m_t
\mu}\over{2s_{\beta}}}s_{2t}\right]
\nonumber\\
&&\qquad\qquad\qquad\qquad\qquad\times
\left[m_Z^2s_{\beta}(\half s_t^2+e_ts_W^2c_{2t})+{{m_t
A_U}\over{2s_{\beta}}}s_{2t}-{{m_t^2}\over{s_{\beta}}}
\right]B_0^{p\tilde t_2\tilde t_2}
\nonumber\\
&-&\left[m_Z^2c_{\beta}(\half c_b^2+e_bs_W^2c_{2b})-{{m_b
A_D}\over{2c_{\beta}}}s_{2b}-{{m_b^2}\over{c_{\beta}}}
\right]
\nonumber\\
&&\qquad\qquad\qquad\qquad\qquad\times
\left[m_Z^2s_{\beta}(\half c_b^2+e_bs_W^2c_{2b})-{{m_b
\mu}\over{2c_{\beta}}}s_{2b}
\right]B_0^{p\tilde b_1\tilde b_1}
\nonumber\\
&-&2\left[m_Z^2c_{\beta}(\quarter+e_bs_W^2)s_{2b}+{{m_b
A_D}\over{2c_{\beta}}}c_{2b}\right]
\left[m_Z^2s_{\beta}(\quarter+e_bs_W^2)s_{2b}+{{m_b
\mu}\over{2c_{\beta}}}c_{2b}\right]B_0^{p\tilde b_1
\tilde b_2}
\nonumber\\
&+&\left[m_Z^2c_{\beta}(\half s_b^2-e_bs_W^2c_{2b})+{{m_b
A_D}\over{2c_{\beta}}}s_{2b}-{{m_b^2}\over{c_{\beta}}}
\right]
\nonumber\\
&&\qquad\qquad\qquad\qquad\qquad\times
\left[m_Z^2s_{\beta}(-\half s_b^2+e_bs_W^2c_{2b})-{{m_b
\mu}\over{2c_{\beta}}}s_{2b}
\right]B_0^{p\tilde b_2\tilde b_2}\Bigg\}\,.
\label{eq:chiabstb}
\end{eqnarray}
And this completes the formulas we need to calculate the momentum dependent
mixing angle $\alpha(p^2)$ and the Higgs masses $m_h$ and $m_H$ in the 
approximation where only top and bottom quarks and squarks are 
considered in the loops.

\vskip.5cm
\section*{Figure Captions}

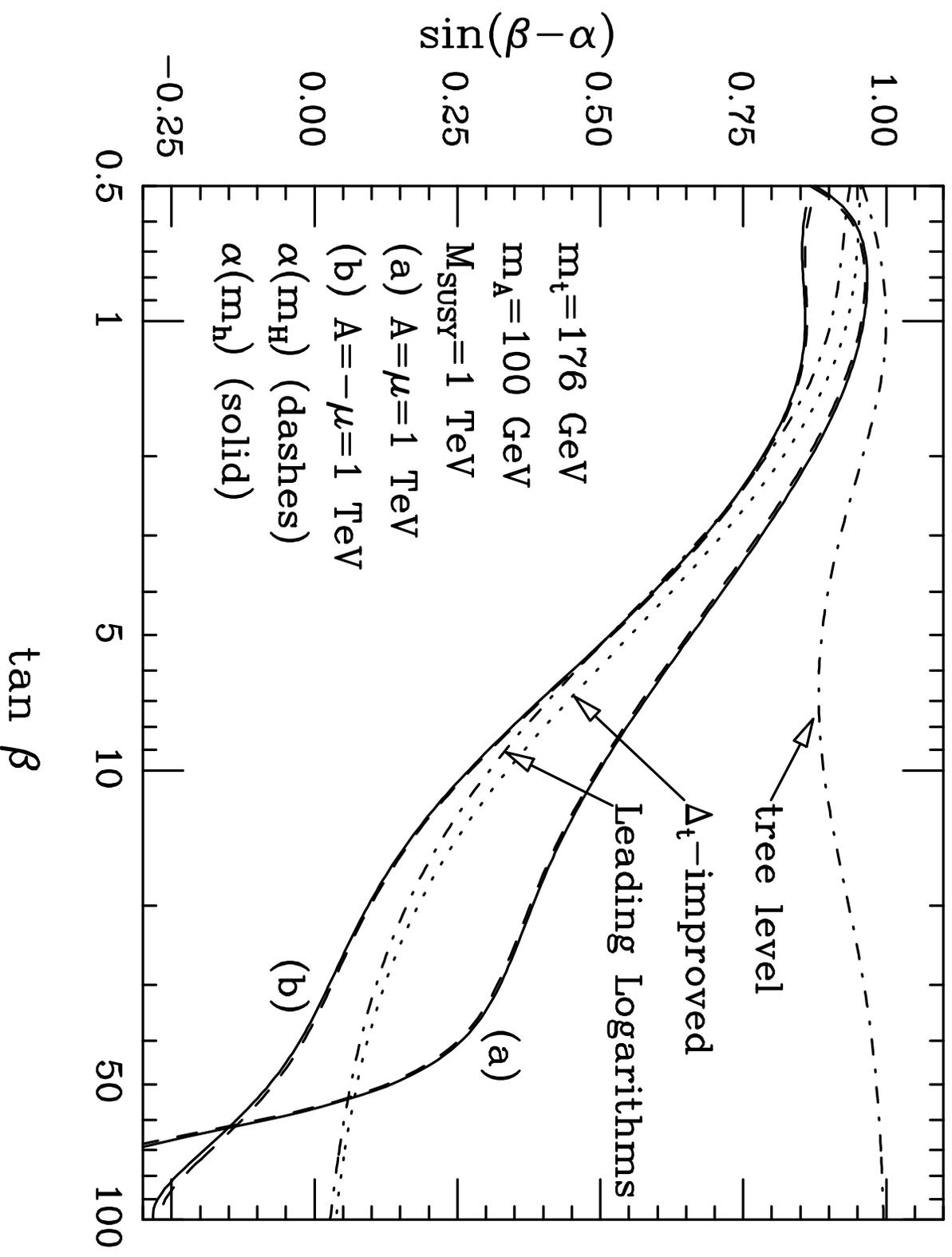
\noindent {\bf Figure 1:} 
Coupling $hZZ$ relative to the SM coupling $H_{SM}ZZ$, 
$\sin(\beta-\alpha)$, as a function of $\tan\beta$.
It is plotted the tree level (upper dot-dashes), the
$\Delta_t$--improved tree level (dots), the leading logarithms
(lower dot-dashes), and the parameter calculated with the 
momentum dependent mixing angle $\alpha(p^2)$. In the last case,
we use the mixing angle evaluated at the heavy Higgs mass scale
$\alpha(m_H)$ (dashes) and the light Higgs mass scale $\alpha(m_h)$
(solid), for two different choices of squark mixing: 
(a) $A=\mu=1$ TeV and (b) $A=-\mu=1$ TeV.

\vskip .4cm
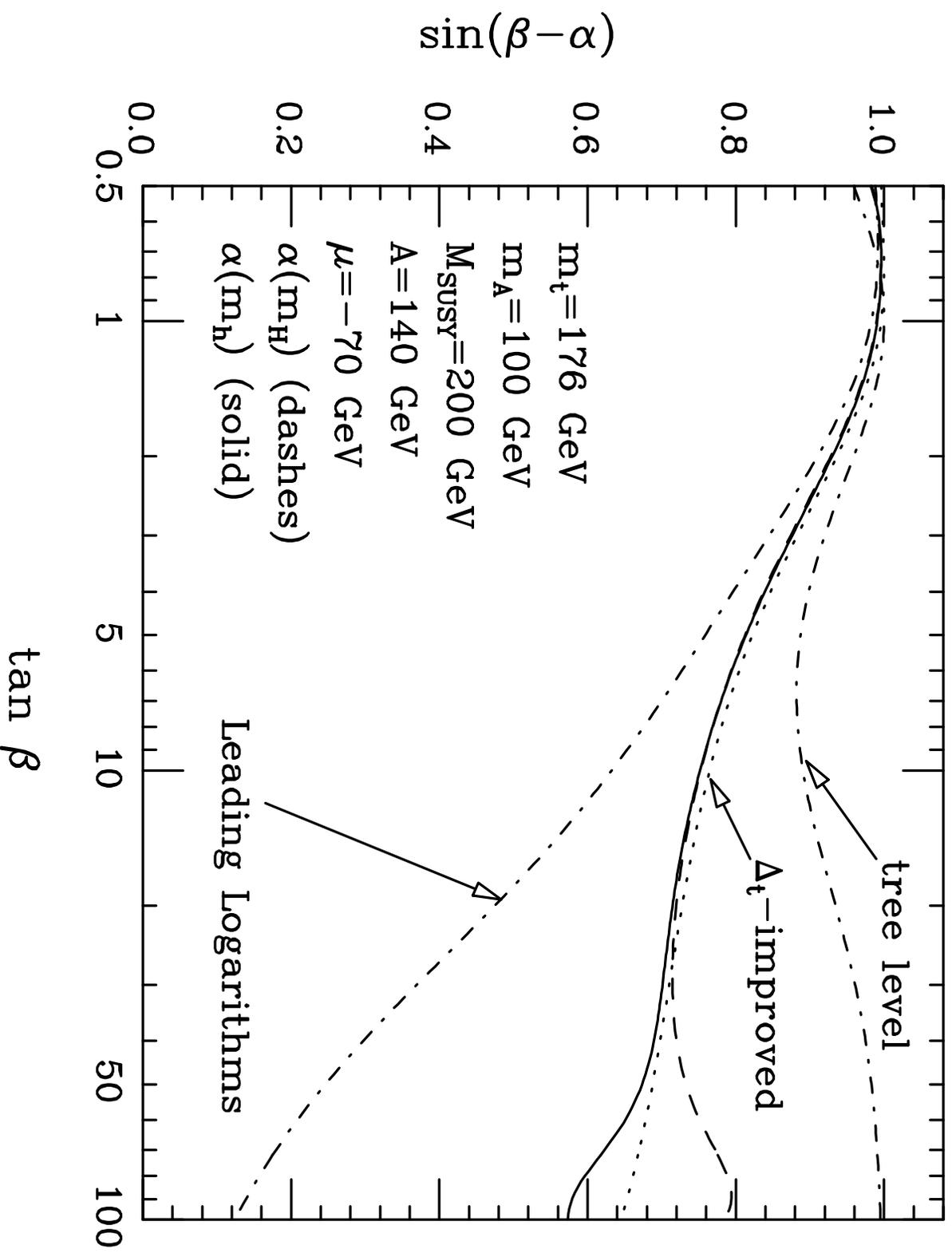
\noindent {\bf Figure 2:} 
Coupling $hZZ$ relative to the SM coupling $H_{SM}ZZ$, 
$\sin(\beta-\alpha)$, as a function of $\tan\beta$.
It is plotted the tree level (upper dot-dashes), the
$\Delta_t$--improved tree level (dots), the leading logarithms
(lower dot-dashes), and the parameter calculated with the 
momentum dependent mixing angle $\alpha(p^2)$. In the last case,
we use the mixing angle evaluated at the heavy Higgs mass scale
$\alpha(m_H)$ (dashes) and the light Higgs mass scale $\alpha(m_h)$
(solid). We take $m_A=100$ GeV, $M_{SUSY}=200$ GeV, $A=140$ GeV,
and $\mu=-70$ GeV.

\vskip .4cm
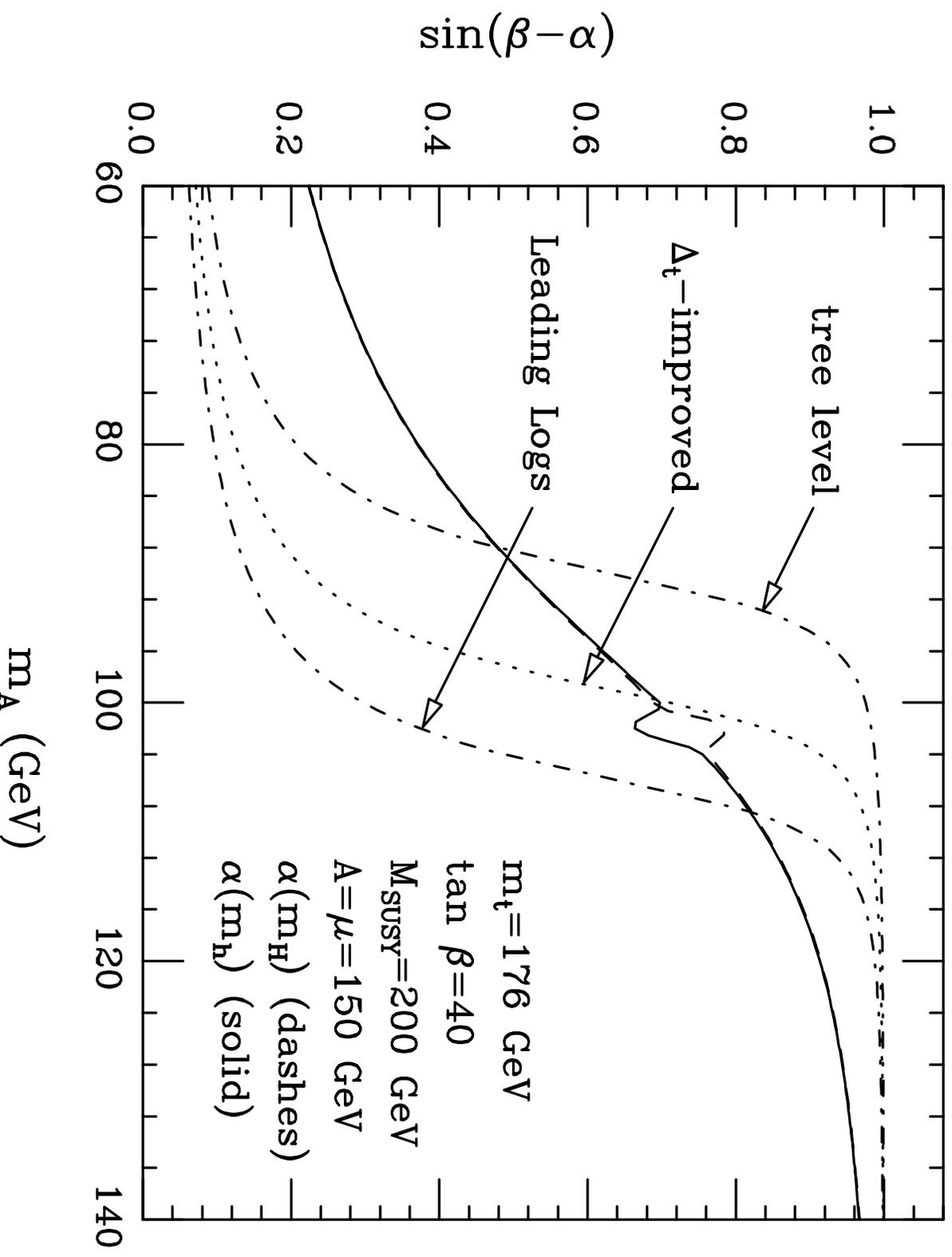
\noindent {\bf Figure 3:}
Coupling $hZZ$ relative to the SM coupling $H_{SM}ZZ$, 
$\sin(\beta-\alpha)$, as a function of $m_A$.
It is plotted the tree level (upper dot-dashes), the
$\Delta_t$--improved tree level (dots), the leading logarithms
(lower dot-dashes), and the parameter calculated with the 
momentum dependent mixing angle $\alpha(p^2)$. In the last case,
we use the mixing angle evaluated at the heavy Higgs mass scale
$\alpha(m_H)$ (dashes) and the light Higgs mass scale $\alpha(m_h)$
(solid), for a squark mixing given by $A=\mu=150$ GeV.

\vskip .4cm
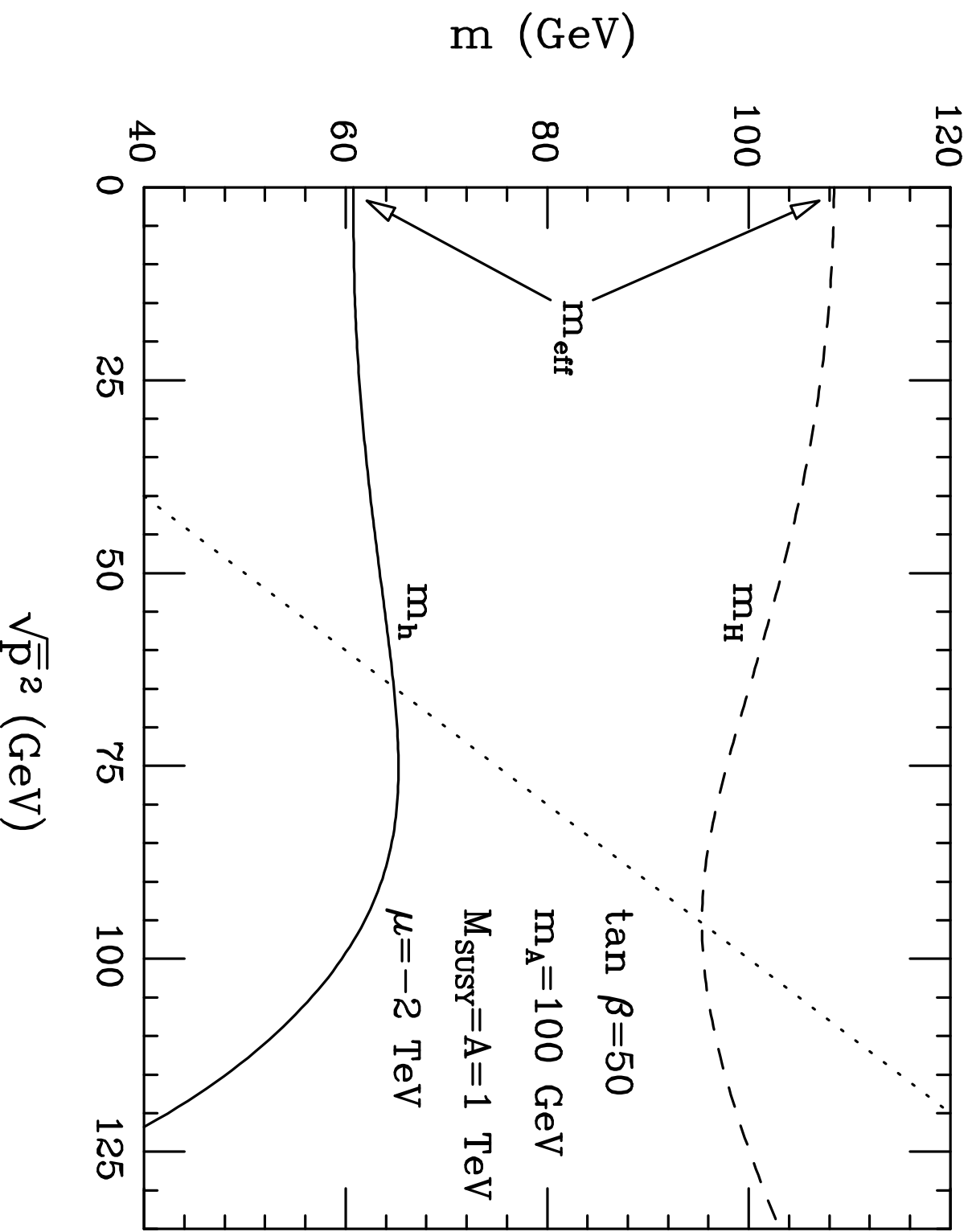
\noindent {\bf Figure 4:}
Neutral Higgs masses $m_h$ (solid) and $m_H$ (dashes) as a function of
the squared root of the argument $\overline{p}^2$ of the self energies. 
We take $\tan\beta=50$, $m_A=100$ GeV, $M_Q=M_U=M_D\equiv M_{SUSY}=1$ TeV,
$A_U=A_D\equiv A=1$ TeV, and $\mu=-2$ TeV. The dotted line is the
diagonal where $m=\sqrt{\overline{p}^2}$, and the intersection of this 
diagonal with the solid (dashed) curve gives us the pole mass $m_h$ 
($m_H$) in our approach. The intersection of the solid (dashed) curve 
with the vertical line defined by $\overline{p}^2=0$ correspond to the 
Higgs mass $m_h$ ($m_H$) calculated with the second derivative of the 
effective potential.

\end{document}